\documentclass[12pt]{article}

\usepackage{amsmath,amsfonts,enumerate,array}
\usepackage{graphicx}
\usepackage[usenames]{color}
\usepackage[english]{babel}
\usepackage{fancybox}
\usepackage{chngpage} 

\newcommand{\captionfonts}{\small}

\makeatletter  
\long\def\@makecaption#1#2{%
  \vskip\abovecaptionskip
  \sbox\@tempboxa{{\captionfonts #1: #2}}%
 \ifdim \wd\@tempboxa >\hsize
    {\captionfonts #1: #2\par}
  \else
    \hbox to\hsize{\hfil\box\@tempboxa\hfil}%
  \fi
  \vskip\belowcaptionskip}
\makeatother   

\usepackage{mciteplus}

\usepackage[colorlinks=true,      linkcolor=blue,      urlcolor=blue,      filecolor=blue,      citecolor=blue,
      pdfstartview=FitH,     pdfpagemode=UseNone,      bookmarksopen=true]{hyperref}  
      
\usepackage[all]{hypcap}     

\begin{document}

\numberwithin{equation}{section}


\mathchardef\mhyphen="2D


\makeatletter
\newcommand{\contraction}[5][1ex]{%
  \mathchoice
    {\contraction@\displaystyle{#2}{#3}{#4}{#5}{#1}}%
    {\contraction@\textstyle{#2}{#3}{#4}{#5}{#1}}%
    {\contraction@\scriptstyle{#2}{#3}{#4}{#5}{#1}}%
    {\contraction@\scriptscriptstyle{#2}{#3}{#4}{#5}{#1}}}%
\newcommand{\contraction@}[6]{%
  \setbox0=\hbox{$#1#2$}%
  \setbox2=\hbox{$#1#3$}%
  \setbox4=\hbox{$#1#4$}%
  \setbox6=\hbox{$#1#5$}%
  \dimen0=\wd2%
  \advance\dimen0 by \wd6%
  \divide\dimen0 by 2%
  \advance\dimen0 by \wd4%
  \vbox{%
    \hbox to 0pt{%
      \kern \wd0%
      \kern 0.5\wd2%
      \contraction@@{\dimen0}{#6}%
      \hss}%
    \vskip 0.2ex%
    \vskip\ht2}}
\newcommand{\contracted}[5][1ex]{%
  \contraction[#1]{#2}{#3}{#4}{#5}\ensuremath{#2#3#4#5}}
\newcommand{\contraction@@}[3][0.06em]{%
  \hbox{%
    \vrule width #1 height 0pt depth #3%
    \vrule width #2 height 0pt depth #1%
    \vrule width #1 height 0pt depth #3%
    \relax}}
\makeatother

\newcommand{\be}{\begin{equation}} 
\newcommand{\ee}{\end{equation}} 
\newcommand{\bea}{\begin{eqnarray}\displaystyle}
\newcommand{\eea}{\end{eqnarray}}
\newcommand{\bt}{\begin{tabular}}
\newcommand{\et}{\end{tabular}}
\newcommand{\bs}{\begin{split}}
\newcommand{\es}{\end{split}}

\newcommand{\nsnsket}{|0_{NS}\rangle^{(1)}\newotimes |0_{NS}\rangle^{(2)}}					
\newcommand{\nsnsbra}{{}^{(1)}\langle 0_{NS}| \newotimes {}^{(2)}\langle 0_{NS}|}		
\newcommand{\nsket}{|0_{NS}\rangle}																									
\newcommand{\nsbra}{\langle 0_{NS}|}																								
\newcommand{\nstket}{|0_{NS}\rangle_t}																										
\newcommand{\nstbra}{{}_t\langle 0_{NS}|}																								
\newcommand{\rmmket}{|0_R^-\rangle^{(1)}\newotimes |0_R^-\rangle^{(2)}}							
\newcommand{\rmmbra}{{}^{(1)}\langle 0_{R,-}| \newotimes {}^{(2)}\langle 0_{R,-}|}	
\newcommand{\rmket}{|0_R^-\rangle}																									
\newcommand{\rmbra}{\langle 0_{R,-}|}																								
\newcommand{\rmtket}{|0_R^-\rangle_t}																								
\newcommand{\rmtbra}{{}_t\langle 0_{R,-}|}																					
\newcommand{\rppket}{|0_R^+\rangle^{(1)}\newotimes |0_R^+\rangle^{(2)}}							
\newcommand{\rppbra}{{}^{(1)}\langle 0_{R,+}| \newotimes {}^{(2)}\langle 0_{R,+}|}	
\newcommand{\rpket}{|0_R^+\rangle}																									
\newcommand{\rpbra}{\langle 0_{R,+}|}																								
\newcommand{\rptket}{|0_R^+\rangle_t}																								
\newcommand{\rptbra}{{}_t\langle 0_{R,+}|}																					
\newcommand{\rpmket}{| 0_R^+\rangle^{(1)} \newotimes | 0_R^-\rangle^{(2)}}					
\newcommand{\rpmbra}{{}^{(1)}\langle 0_{R,+}| \newotimes {}^{(2)}\langle 0_{R,-}|}	
\newcommand{\rmpket}{| 0_R^-\rangle^{(1)} \newotimes | 0_R^+\rangle^{(2)}}					
\newcommand{\rmpbra}{{}^{(1)}\langle 0_{R,-}| \newotimes {}^{(2)}\langle 0_{R,+}|}	

\newcommand{\nsutvket}{|0_{NS}\rangle^{(1)}\otimes |0_{NS}\rangle^{(2)}}
\newcommand{\nsutvbra}{{}^{(1)}\langle 0_{NS}| \otimes {}^{(2)}\langle 0_{NS}|}
\newcommand{\nstvket}{|0_{NS}\rangle}
\newcommand{\nstclose}{0_{NS}\rangle}
\newcommand{\nstvbra}{\langle 0_{NS}|}
\newcommand{\nstpket}{|0\rangle_t}
\newcommand{\nstpbra}{{}_t\langle 0|}
\newcommand{\rmutvket}{|0_R^-\rangle^{(1)}\otimes |0_R^-\rangle^{(2)}}
\newcommand{\rmutvbra}{{}^{(1)}\langle 0_{R,-}| \otimes {}^{(2)}\langle 0_{R,-}|} 
\newcommand{\rmtvket}{|0_R^-\rangle}
\newcommand{\rmtvbra}{\langle 0_{R,-}|}
\newcommand{\rmtpket}{|0_R^-\rangle_t}
\newcommand{\rmtpbra}{{}_t\langle 0_{R,-}|}
\newcommand{\rputvket}{|0_R^+\rangle^{(1)}\otimes |0_R^+\rangle^{(2)}}
\newcommand{\rputvbra}{{}^{(1)}\langle 0_{R,+}| \otimes {}^{(2)}\langle 0_{R,+}|} 
\newcommand{\rptvket}{|0_R^+\rangle}
\newcommand{\rptvbra}{\langle 0_{R,+}|}
\newcommand{\rptpket}{|0_R^+\rangle_t}
\newcommand{\rptpbra}{{}_t\langle 0_{R,+}|}
\newcommand{\stp}{\sigma_2^+}
\newcommand{\stm}{\sigma_2^-}

\renewcommand{\a}{\alpha}	
\renewcommand{\b}{\beta}
\newcommand{\g}{\gamma}		
\newcommand{\G}{\Gamma}
\renewcommand{\d}{\delta}
\newcommand{\D}{\Delta}
\renewcommand{\c}{\chi}			
\newcommand{\C}{\Chi}
\newcommand{\p}{\psi}			
\renewcommand{\P}{\Psi}
\newcommand{\s}{\sigma}		
\renewcommand{\S}{\Sigma}
\renewcommand{\t}{\tau}		
\newcommand{\e}{\epsilon}
\newcommand{\n}{\nu}
\newcommand{\m}{\mu}
\renewcommand{\r}{\rho}
\renewcommand{\l}{\lambda}

\newcommand{\nn}{\nonumber\\} 		
\newcommand{\newotimes}{}  				
\newcommand{\diff}{\,\text{d}}		
\newcommand{\h}{{1\over2}}				
\newcommand{\Gf}[1]{\G \Big{(} #1 \Big{)}}	
\newcommand{\floor}[1]{\left\lfloor #1 \right\rfloor}
\newcommand{\ceil}[1]{\left\lceil #1 \right\rceil}

\def\cA{{\cal A}} \def\cB{{\cal B}} \def\cC{{\cal C}}
\def\cD{{\cal D}} \def\cE{{\cal E}} \def\cF{{\cal F}}
\def\cG{{\cal G}} \def\cH{{\cal H}} \def\cI{{\cal I}}
\def\cJ{{\cal J}} \def\cK{{\cal K}} \def\cL{{\cal L}}
\def\cM{{\cal M}} \def\cN{{\cal N}} \def\cO{{\cal O}}
\def\cP{{\cal P}} \def\cQ{{\cal Q}} \def\cR{{\cal R}}
\def\cS{{\cal S}} \def\cT{{\cal T}} \def\cU{{\cal U}}
\def\cV{{\cal V}} \def\cW{{\cal W}} \def\cX{{\cal X}}
\def\cY{{\cal Y}} \def\cZ{{\cal Z}}

\def\mC{\mathbb{C}} \def\mP{\mathbb{P}}  
\def\mR{\mathbb{R}} \def\mZ{\mathbb{Z}} 
\def\mT{\mathbb{T}} \def\mN{\mathbb{N}}
\def\mH{\mathbb{H}} \def\mX{\mathbb{X}}
\def\CP{\mathbb{CP}}
\def\RP{\mathbb{RP}}
\def\Z{\mathbb{Z}}
\def\N{\mathbb{N}}
\def\H{\mathbb{H}}

\newcommand{\Zd}{\ensuremath{ Z^{\dagger}}}
\newcommand{\Xd}{\ensuremath{ X^{\dagger}}}
\newcommand{\Ad}{\ensuremath{ A^{\dagger}}}
\newcommand{\Bd}{\ensuremath{ B^{\dagger}}}
\newcommand{\Ud}{\ensuremath{ U^{\dagger}}}
\newcommand{\Td}{\ensuremath{ T^{\dagger}}}
\newcommand{\T}[3]{\ensuremath{ #1{}^{#2}_{\phantom{#2} \! #3}}}		
\newcommand{\tr}{\operatorname{tr}}
\newcommand{\sech}{\operatorname{sech}}
\newcommand{\Spin}{\operatorname{Spin}}
\newcommand{\Sym}{\operatorname{Sym}}
\newcommand{\Com}{\operatorname{Com}}
\def\adj{\textrm{adj}}
\def\id{\textrm{id}}
\def\pb{\ov\psi}
\def\pt{\widetilde{\psi}}
\def\at{\widetilde{\a}}
\def\cb{\ov\chi}
\def\db{\bar\partial}
\def\delb{\bar\partial}
\def\dbar{\ov\partial}
\def\dag{\dagger}
\def\dalpha{{\dot\alpha}}
\def\dbeta{{\dot\beta}}
\def\dgamma{{\dot\gamma}}
\def\ddelta{{\dot\delta}}
\def\ad{{\dot\alpha}}
\def\bd{{\dot\beta}}
\def\dg{{\dot\gamma}}
\def\dd{{\dot\delta}}
\def\th{\theta}
\def\Th{\Theta}
\def\eb{{\ov \epsilon}}
\def\gb{{\ov \gamma}}
\def\wb{{\ov w}}
\def\Wb{{\ov W}}
\def\D{\Delta}
\def\DD{\Delta^\dag}
\def\Db{\ov D}
\def\ov{\overline}
\def\Slash{\, / \! \! \! \!}
\def\dslash{\partial\!\!\!/} 
\def\Dslash{D\!\!\!\!/\,\,}
\def\fslash#1{\slash\!\!\!#1}
\def\Fslash#1{\slash\!\!\!\!#1}
\def\del{\partial}
\def\delb{\bar\partial}
\newcommand{\ex}[1]{{\rm e}^{#1}} 
\def\ii{{i}}
\newcommand{\vs}[1]{\vspace{#1 mm}}
\newcommand{\ve}{{\vec{\e}}}
\newcommand{\shalf}{\frac{1}{2}}
\newcommand{\lb}{\rangle}
\newcommand{\al}{\ensuremath{\alpha'}}
\newcommand{\ap}{\ensuremath{\alpha'}}
\newcommand{\ft}[2]{{\textstyle {\frac{#1}{#2}} }}

\newcommand{\rmd}{\mathrm{d}}
\newcommand{\rmx}{\mathrm{x}}
\def\tA{ {\widetilde A} } 
\def\one{{\hbox{\kern+.5mm 1\kern-.8mm l}}}
\def\zero{{\hbox{0\kern-1.5mm 0}}}
\def\eq#1{(\ref{#1})}
\newcommand{\secn}[1]{Section~\ref{#1}}
\newcommand{\tbl}[1]{Table~\ref{#1}}
\newcommand{\fig}{Fig.~\ref}
\def\sqi{{1\over \sqrt{2}}}
\newcommand{\hsp}{\hspace{0.5cm}}
\def\half{{\textstyle{1\over2}}}
\let\ci=\cite \let\re=\ref
\let\se=\section \let\sse=\subsection \let\ssse=\subsubsection
\newcommand{\dpb}{D$p$-brane}
\newcommand{\dpbs}{D$p$-branes}
\def\gh{{\rm gh}}
\def\sgh{{\rm sgh}}
\def\NS{{\rm NS}}
\def\R{{\rm R}}
\def\Qp{Q_{\rm P}}
\def\QP{Q_{\rm P}}
\newcommand\dott[2]{#1 \! \cdot \! #2}
\def\eo{\overline{e}}
\newcommand{\bb}{\bigskip}
\newcommand{\ac}[2]{\ensuremath{\{ #1, #2 \}}}
\renewcommand{\ell}{l}
\newcommand{\z}{\ell}
\newcommand{\bm}{\bibitem}
\newcommand\com[2]{[#1,\,#2]}

\newcommand{\bra}[1]{{\langle {#1} |\,}}
\newcommand{\ket}[1]{{\,| {#1} \rangle}}
\newcommand{\braket}[2]{\ensuremath{\langle #1 | #2 \rangle}}
\newcommand{\Braket}[2]{\ensuremath{\langle\, #1 \,|\, #2 \,\rangle}}
\newcommand{\norm}[1]{\ensuremath{\left\| #1 \right\|}}
\def\corr#1{\left\langle \, #1 \, \right\rangle}
\def\vac{|0\rangle}


\vspace{16mm}

 \begin{center}
{\LARGE Second order effect of twist deformations }\\
\vspace {2mm}
   {\LARGE in the D1D5 CFT}

\vspace{18mm}
{\bf  Zaq Carson\footnote{carson.231@osu.edu}, Shaun Hampton\footnote{hampton.197@osu.edu}, and Samir D. Mathur\footnote{mathur.16@osu.edu}
\\}
\vspace{15mm}
Department of Physics,\\ The Ohio State University,\\ Columbus,
OH 43210, USA\\ 
\vspace{8mm}
\end{center}

\vspace{10mm}

\thispagestyle{empty}
\begin{abstract}

\vspace{3mm}

Thermalization in the D1D5 CFT should occur via interactions caused by the twist operator, which deforms the theory off its free orbifold point. Earlier studies investigating this deformation at first order did not show any definite evidence of thermalization. In this paper we study the deformation to second order, where we do expect to see the effects that should give thermalization. We compute the effect of two twist operators on an initial vacuum state, which generates a squeezed state analogous to the case for a single twist.  We obtain expressions for  the Bogoliubov coefficients in this 2-twist case.

\end{abstract}
\newpage

\section{Introduction}
\label{intr}\setcounter{footnote}{0}

What happens when matter collapses to make a black hole? The details of this process are unclear, and potentially of great importance in obtaining a full understanding of quantum gravity. In string theory, this gravitational process can be put in a setting where we have a CFT dual \cite{adscft}, and we can try to study the collapse process in its dual field theory description. 

The CFT dual is itself  difficult  to study, since the gravitational theory is dual to a strongly coupled CFT. Remarkably, one obtains many interesting results about black holes by using the CFT at its `free'  or `orbifold' point \cite{sv,lm1, lm2,orbifold1,orbifold2,deformation,orbifold3}. At this coupling the CFT is given by several symmetrized copies of a free CFT. This symmetrizaton introduces twist operators $\sigma_k$ into the theory. The free theory reproduces the entropy and greybody factors of near-extremal black holes \cite{radiation}, but it cannot describe their formation. The reason is that black hole formation is expected to be dual to a thermalization in the CFT, and in a free theory an initial excitation will not in general thermalize. 

We are therefore led to explore the deformation of the theory away from the orbifold point. This deformation is given by the operator
 \be
\hat O_{\dot A\dot B}(w_0,\bar{w}_0)=
\left[{1\over 2\pi i} \int\limits_{w_0} dw G^-_{\dot A} (w)\right]
\left[{1\over 2\pi i} \int\limits_{\bar w_0} d\bar w 
\bar G^-_{\dot B} (\bar w)\right]
\sigma_2^{++}(w_0,\bar{w}_0)
\ee
The index notations are explained in Appendix \ref{ap:CFT-notation}, but we note here the general structure of this operator. There is a twist $\sigma_2$, which takes two copies of the free CFT and joins them together. If these copies were defined on circles of length $2\pi R$, the twist will give one copy of the CFT on a circle of length $4\pi R$. In addition to this twist we have the application of left and right supercharges $G$.

As explained in \cite{acm1,acm2}, the $G$ operators can be removed from the $\sigma_2$ by writing them as contours surrounding the $\sigma_2$, and then stretching these contours away till they act on the initial or final states of the process. Thus a core part of the effect of the deformation operator is given by the action of the `bare twist' $\sigma_2$. In \cite{acm2} it was found that if we apply one operator $\sigma_2(w)$ to the vacuum $|0\rangle$, the resulting state is  a squeezed state, with the schematic form
\be
e^{\gamma^B_{mn} \alpha_{-m}\alpha_{-n}+ \gamma^F_{rs} d_{-r}d_{-s}}|0\rangle
\label{one}
\ee

The coefficients $\gamma^B, \gamma^F$ were computed in \cite{acm2} for the case where the initial copies of the CFT lived on singly twisted circles. In \cite{peet3} they were computed by an alternative method that made their basic properties manifest.  In \cite{cmt} they were computed for the case where the initial strings had arbitrary order twists $M,N$, but where $m,n\gg 1$. In \cite{chmt1,chmt2} they were computed for arbitary $M,N$ and arbitary $m,n$. In \cite{peet1,peet2} the mixing of operators under the twist deformations was analyzed.  All of these computations addressed the case of a single $\sigma_2$ insertion, but it was not clear that thermalization can be observed at this order.

In the present paper we perform a 1-loop computation, where we have {\it two} twist insertions: the first twist joins two untwisted copies of the CFT to a copy living on  a double circle, and the second twist returns us to the two untwisted copies. We argue that starting from the vacuum state $|0\rangle$ we again get a squeezed state of the form (\ref{one}), and we compute the required $\gamma^B, \gamma^F$. We find that these coefficients can be expressed as a finite sum over a bilinear product of hypergeometric functions. 

One apparently puzzling feature of our computations is the role of sectors with new periodicities for fermions. We start with both copies of the CFT in the Ramond (R) sector. After the application of both $\sigma_2$ factors we are left with a contribution where both CFTs are in the R sector, but we can also obtain a nonzero result for the amplitude if we let both the final copies of the CFT be in the NS sector.  As we will explain later, we believe that the NS sectors should not be included as possible final states. We still compute the amplitude to transition to these sectors, but we will drop this contribution from our final analysis. 

The plan of this paper is as follows.  In Section \ref{Orbifold Point}, we introduce the orbifold CFT.  In Section \ref{Outline}, we lay out the calculation for determining the state $|\chi(w_1,w_2)\rangle$.  This state contains three major components.  In Section \ref{Boson} we calculate the coefficients in the bosonic component.  In Section \ref{sec:f} we calculate the coefficients for the fermionic component in the NS sector.  In Section \ref{Ramond} we calculate the coefficients for the fermionic component in the Ramond sector.  In Section \ref{NS-R Sectors}, we argue that the contribution from the NS sector is nonphysical.  In Section \ref{Analysis}, we perform some numerical analyses on the R sector results to obtain some simplifying limits.

\section{The D1D5 CFT at the orbifold point}\label{Orbifold Point}

In this section we summarize some properties of the D1D5 CFT at the orbifold point and the deformation operator that we will use to perturb away from the orbifold point. For more details, see \cite{acm2}.

\subsection{The D1D5 CFT}

Consider type IIB string theory, compactified as
\be
M_{9,1}\rightarrow M_{4,1}\times S^1\times T^4.
\label{compact}
\ee
Wrap $N_1$ D1 branes on $S^1$, and $N_5$ D5 branes on $S^1\times
T^4$. We think of the $S^1$ as being large compared to the $T^4$, so
that at low energies we look for excitations only in the direction
$S^1$.  This low energy limit gives a conformal field theory (CFT) on
the circle $S^1$.

We can now vary the moduli of string theory (the string coupling $g$, the
shape and size of the torus, the values of flat connections for gauge
fields etc.). These changes move us to different points in the moduli
space of the CFT. It has been conjectured that we can move to a point
called the `orbifold point' where the CFT is particularly simple
\cite{orbifold2}. At this orbifold point the CFT is
a 1+1 dimensional sigma model. We will work in the Euclidean theory, where
the base space is a cylinder spanned by the coordinates 
\be
\tau, \sigma: ~~~0\le \sigma<2\pi, ~~~-\infty<\tau<\infty
\ee
The target space of this sigma model is the symmetrized product of
$N_1N_5$ copies of $T^4$,
\be
(T_4)^{N_1N_5}/S_{N_1N_5},
\ee
with each copy of $T^4$ giving 4 bosonic excitations $X^1, X^2, X^3,
X^4$. It also gives 4 fermionic excitations, which we call $\psi^1,
\psi^2, \psi^3, \psi^4$ for the left movers, and $\bar\psi^1,
\bar\psi^2,\bar\psi^3,\bar\psi^4$ for the right movers.  The central charge of the theory with fields
$X^i, \psi^i, ~i=1\dots 4$ is
$c=6$. The total central charge of the entire system is thus $6 N_1N_5$.

In \cite{acm2} it was noted that we can write the deformation operator as
\be
\hat O_{\dot A\dot B}(w_0)=\Big [{1\over 2\pi i} \int _{w_0} dw G^-_{\dot A} (w)\Big ]\Big [{1\over 2\pi i} \int _{\bar w_0} d\bar w \bar G^-_{\dot B} (\bar w)\Big ]\sigma_2^{++}(w_0).
\label{pert}
\ee
We will, however, work only with the left-moving part of (\ref{pert}).  The right-moving sector is fully analogous.

\subsection{NS and R vacua}
Consider a single copy of the $c=6$ CFT. The lowest energy state of the left-moving sector is the NS vacuum, 
\be
|0_{NS}\rangle \,, \qquad\quad h = 0, \quad m = 0 \,
\ee
where $h$ denotes the $L_0$ eigenvalue.
However, we will for the most part study the CFT in the R sector. In particular we are interested in the R vacua denoted by\footnote{There are two other left-moving R ground states, which may be obtained by acting with fermion zero modes. Including right-movers then gives a total of 16 R-R ground states.  These R-R ground states are described in detail in Appendix \ref{RVN}}
\be
|0_{R}^{\pm}\rangle \,, 
\qquad\quad h=\frac14, \quad m=\pm\frac12 \,.
\ee
We can relate the NS and R sectors using spectral flow \cite{spectralref}.  In particular, spectral flow by $\a = \pm 1$ in the left-moving sector produces the transformations
\bea
\alpha= 1 : && \ket{0_R^-} \rightarrow \ket{0_{NS}} \,, 
\quad \ket{0_{NS}}  \rightarrow \ket{0_R^+}  \cr
\alpha= -1 : && \ket{0_R^+} \rightarrow \ket{0_{NS}} \,, 
\quad \ket{0_{NS}}  \rightarrow \ket{0_R^-}.  \label{eq:sfstates}
\eea

\section{Outline of the calculations}\label{Outline}

The deformation operator is composed of a supercharge contour acting on a twist operator. The action of the supercharge contour can be split off from the main computation, which involves the effects of the twist operators. In this paper we will focus of the effect of the twist operators alone. 

Since we are looking at deformations to second order, we will have two twist-2 operators acting on our initial state. If $n$ copies of the CFT are twisted together, we say that we have an $n$-times wound `component string'. We will start with the simplest case of two singly wound component strings. The twist from the first deformation operator will change this to a single component string with winding 2, while the second twist will take this new component string back to two component strings of winding 1 each. Thus we are looking at a `1-loop' process in the interacting CFT. We believe that such second order effects will lead to the thermalization that we seek. 

Each of the initial component strings are taken to be in the negative Ramond vacuum state $\rmket$.  The final state after the application of two twists will be denoted by $|\c(w_1)(w_2)\rangle$. Thus:
\bea
|\c(w_1,w_2)\rangle & \equiv & \s_2^+(w_2)\s_2^+(w_1) \rmmket.
\label{stateq}
\eea
Here $w$ is a coordinate on the cylinder, with
\be
w=\tau+i\sigma,
\ee
where $\sigma$ is an angular coordinate for the compact spacial dimension and $\tau$ is a Euclideanized time coordinate.  In (\ref{stateq}), we assume that $\tau_2 > \tau_1$.

In the region $\t_1 < \t < \t_2$,  the two component strings are joined together to form a doubly wound componnet string. The bosonic fields are now periodic only after an interval $\Delta \sigma = 4\pi$. The fermionic fields, a priori, can be periodic or antiperiodic after $\Delta \sigma = 4\pi$.  Outside the above interval of $\tau$ we have two singly wound component strings.  We will now map this configuration to a double cover of the cylinder.  To do this, we first map the cylinder into the complex plain with coordinate $z$ and then map the complex plane to a double cover of itself.

\begin{figure}[tbh]
\begin{center}
\includegraphics[width=0.3\columnwidth]{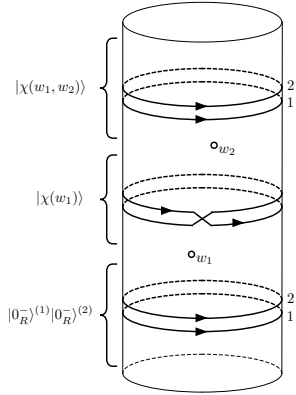}
\end{center}
\caption{The cylinder with twist insertions at $w_1$ and $w_2$.  Below the first twist we have the negative Ramond vacuum on each component string.  Above both twists we have the state $|\chi(w_1,w_2)\rangle$, which we will compute.  In the intermediate regions we have a single doubly-wound component string in the state $|\chi(w_1,w_2)\rangle$.  This state was computed in \cite{acm1} and is not used here.}
\label{figone}
\end{figure}

Around any point $z$ on the plane, a fermionic field can have either periodic or antiperiodic boundary conditions. If we insert only local operators at $z$, then the fermion will be periodic (i.e. it will have NS boundary conditions). If we wish to have instead the anti-periodic boundary condition (i.e., the R boundary condition) then we must insert a spin field $S^\pm$ at the point $z$. The spin field carries  a charge $j=\pm \h$ and has dimension $h={1\over 4}$. 

  The initial state (the `in' state) has two singly wound component strings, each in the R sector. This brings in two spin fields, one from each component string.  We write
\bea
\rmmket = S^{(1)-}(\tau = -\infty)S^{(2)-}(\tau = -\infty)\nsnsket.
\eea
In addition, the twist operator $\s_2^+$ contains a spin field:
\bea
\s_2^+(w) = S^+(w) \s_2(w).
\eea
We will compute an amplitude by taking the inner product of the state we generate after the twists with some state of our choosing. This state with which we `cap' the cylinder may then bring additional spin fields of its own. 

Each of the above mentioned spin fields will have some position in our double cover.  We can remove a spin field by performing a spectral flow around that point, since a spectral flow can map an R sector state to an NS sector state. We will thus perform a series of spectral flows at various points in the double cover in order to remove these spin field insertions.  Once all the spin fields have been removed, the locations where they were inserted will have just the local NS vacuum inserted there, and we can close the corresponding puncture at that location with no insertions. Any contour can then be smoothly deformed through such a location.  It is through these  smooth deformations that we will be able to map in states to out states and determine the nature of $|\c(w_1,w_2)\rangle$

We now divide the remainder of this section into four parts.  In the first part, we outline the coordinate changes used to map the cylinder into a double cover of the complex plane and identify the images of all critical points.  In the second part, we introduce the mode operators on the cylinder.  In the third part, we present the general form of the state $|\c(w_1,w_2)\rangle$ in terms of these mode operators.  This general form motivates capping with certain types of states, which allows us to derive expressions for the parameters of our ansatz.  Finally, we determine the spectral flows needed for each type of capping state based on the images of the spin fields of those states.

\subsection{Coordinate maps}\label{CoordinateMaps}
\begin{figure}[tbh]
\begin{center}
\includegraphics[width=0.4\columnwidth]{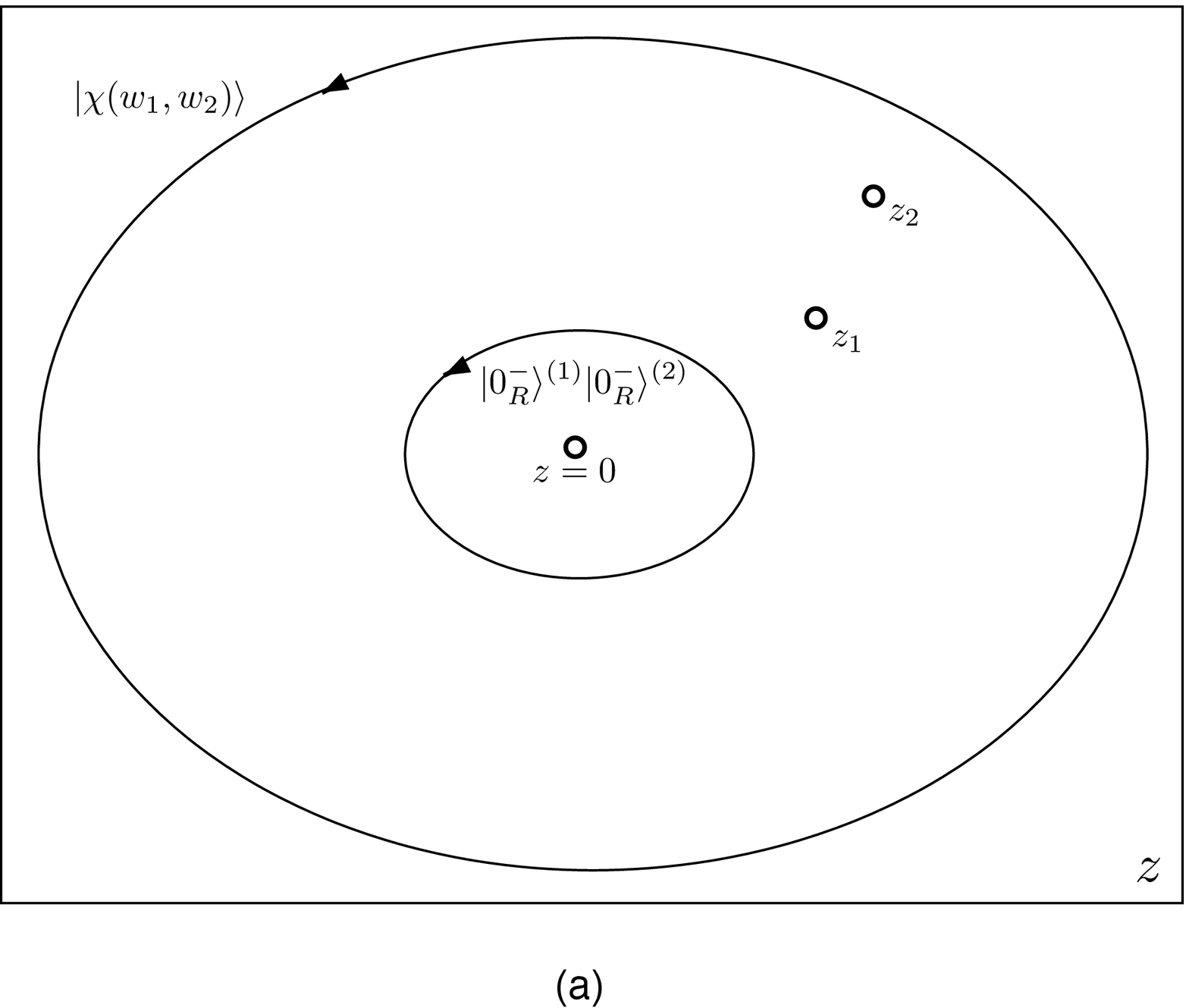} $\qquad\qquad$ \includegraphics[width=0.4\columnwidth]{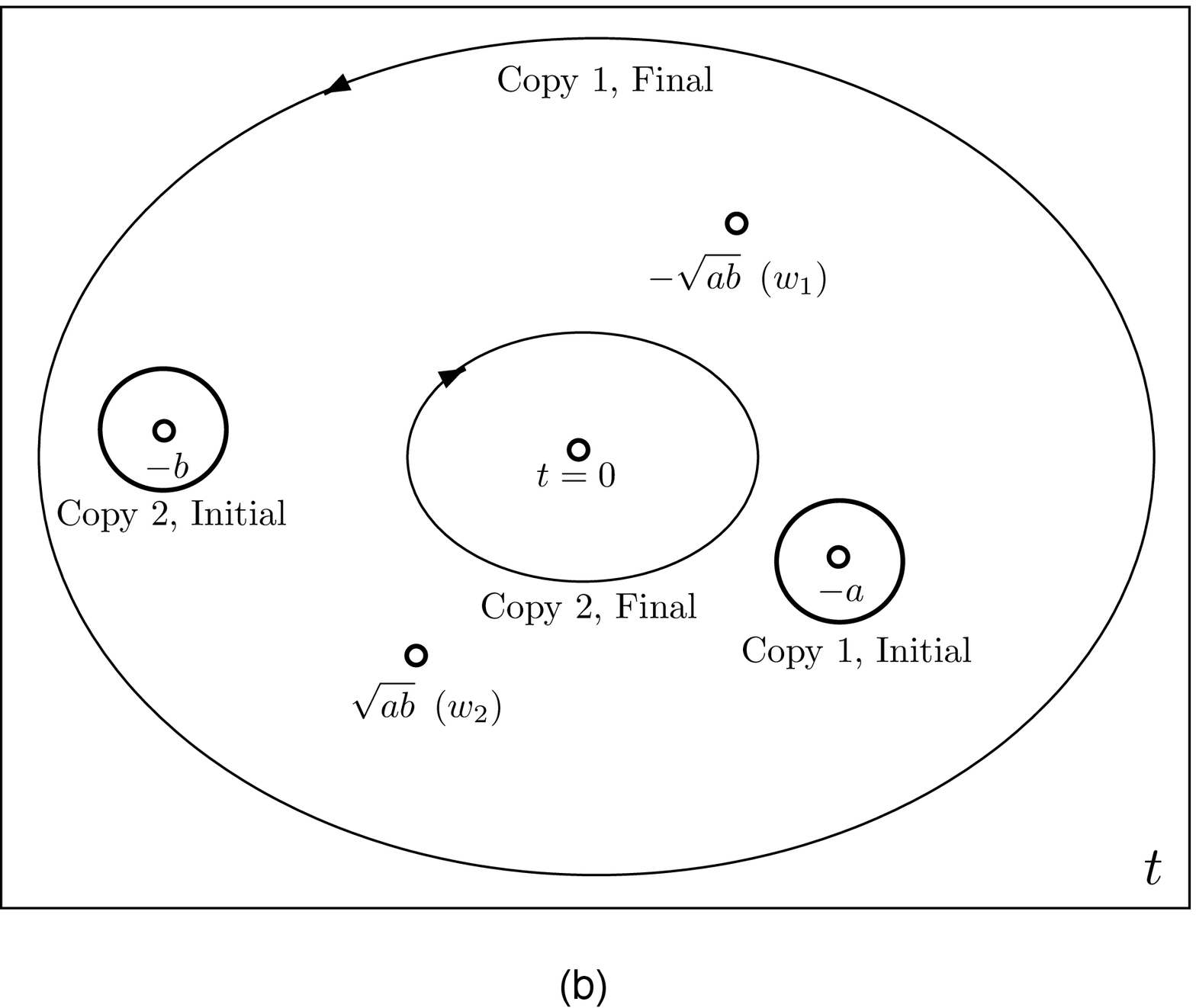}
\end{center}
\caption{The $z$ plane (a) and $t$ plane (b) with all the relevant image points labeled.  The intermediate state $|\chi(w_1)\rangle$ is not depicted.  In the $z$ plane the $\tau$ coordinate maps to the radial coordinate, while the $\s$ coordinate maps to the phase.  In the $t$ plane there are no simple directions corresponding to $\t$ and $\s$.}
\label{CoordinateMapFigure}
\end{figure}

The effects of our coordinate maps are illustrated in Figure \ref{CoordinateMapFigure}.  First we map the cylinder to the complex plane through the map:
\bea
z & = & e^w ~\equiv~ e^{\t + i\s}.
\eea
Here the in states at $\tau = -\infty$ map to the origin of the complex plane, while the out states at $\tau = +\infty$ map to $z = \infty$.  Our fields are still double-valued near these points, as there are two component strings in both the in and out states.  The timelike direction $\tau$ becomes the radial direction, while the spacelike $\s$ becomes the phase of the $z$ plane.  Because $\tau_2 > \tau_1$, we have:
\bea \label{zmap}
|z_2| &=& e^{\t_2} ~>~ |z_1| ~=~ e^{\t_1}.
\eea

Now we map the $z$ plane into a double cover of itself.  We need to ensure that this map will separate out the two distinct copies of our fields for the in and out states.  Thus we require that $z=0$ and $z=\infty$ both have two distinct image points.  It is also useful to ensure that near these points we have leading order relations of the form $z \sim t^{\pm 1}$.  Such a map can in general be written as:
\bea \label{tmap}
z &=& {(t+a)(t+b)\over t}.
\eea
Here, the point $z = \infty$ maps to both $t = \infty$ and $t = 0$.  Near $t=\infty$ we have the leading order behavior $z \sim t$, while near $t=0$ we have the behavior $z \sim t^{-1}$.  The point $z = 0$ also has two images, at $t= -a$ and $t = -b$.  This allows us to split up the images of the two copies for our in and out states, thus leaving us with only single-valued fields in the $t$ plane.

At this point, we have some choice of which component strings map to which regions in the $t$ plane.  The in state has images at $t = -a$ and $t = -b$, but the physics is independent of which component string we call copy 1 and which we call copy 2.  Indeed, the map (\ref{tmap}) is symmetric under the interchange $a \leftrightarrow b$.  Similarly, we may choose which component string of the out state maps to $t = \infty$ and which component string maps to $t = 0$.  Here we present the conventions used in this paper:
\bea
\text{Copy 1 In State} & \to & t = -a \nn
\text{Copy 2 In State} & \to & t = -b \nn
\text{Copy 1 Out State} & \to & t = \infty \nn
\text{Copy 2 Out State} & \to & t = 0.
\eea

We also need to know the images of the twist insertions at $w_1$ and $w_2$.  These insertions create bifurcation points in the double cover, so we find the images by solving for these bifurcation points.
\bea
{\diff z \over \diff t} & = & 1 - {ab \over t^2} ~=~ 0.
\eea
This relation has two solutions, corresponding to our two twist insertions.
\bea \label{twistimages}
t_1 & = & -\sqrt{ab} \nn
t_2 & = & \sqrt{ab}.
\eea
This in turn gives us a relationship between the double cover map parameters $a,b$ and our twist insertion points $w_1,w_2$:
\bea
z_1 & = & a+b - 2\sqrt{ab} ~=~ e^{w_1} \nn
z_2 & = & a+b + 2\sqrt{ab} ~=~ e^{w_2},
\eea
where we choose the branch of $\sqrt{ab}$ so as to maintain $|z_2| > |z_1|$.  As expected, this relationship is unaffected by the interchange of $a$ and $b$.

\subsection{Modes on the cylinder}
In order to determine the final state $|\chi(w_1,w_2)\rangle$ we first introduce the bosonic and fermionic modes that live on the cylinder in the region $\t > \t_2$.  For the bosons, we have:
\bea\label{BosonCylinderMode}
\alpha_{A\dot A, n}^{(i)f} &=& {1\over 2\pi} \int\limits_{\sigma=0}^{2\pi} X_{A\dot A}^{(i)}(w) e^{nw} \diff w,
\eea
where $f$ stands for final modes, indicating that we are working in the region above both twists. The commutation relations are
\bea\label{BosonCommutator}
[\alpha_{A\dot A, n}^{(i)f}, \alpha_{B\dot B, m}^{(j)f}] &=& -n\epsilon_{AB}\epsilon_{\dot A\dot B}\, \delta^{ij} \delta_{n+m,0}\;.
\eea

For fermions, we can have two different types of modes depending on which sector we are in, NS or R.  In the NS sector, the modes are indexed by half-integers $r$:
\bea\label{FermionNSCylinderMode}
d^{(i)f,\a A}_{r} &=& {1\over 2\pi i} \int\limits_{\s=0}^{2\pi} \psi^{(i)\alpha A}(w) e^{rw}\diff w,
 \eea
with anticommutation relations
\bea
\left \{ d^{(i)f,\a A}_{r}, d^{(j)f,\b B}_{s} \right \} &=& -\e^{\a\b}\e^{AB}\d^{(i)(j)}\d_{r+s,0}\;.
\eea
Similarly, the modes in the Ramond sector are:
\bea\label{FermionRCylinderMode}
d^{(i)f,\a A}_{n} &=& {1\over 2\pi i} \int\limits_{\s=0}^{2\pi} \psi^{(i)\alpha A}(w) e^{nw}\diff w,
\eea
with integer $n$.  These modes have anticommutation relations
\bea
\left \{ d^{(i)f,\a A}_{n}, d^{(j)f,\b B}_{m} \right \} &=& -\e^{\a\b}\e^{AB}\d^{(i)(j)}\d_{n+m,0}\;.
\eea

In the Ramond sector there are fermion zero modes.  While bosonic zero modes annihilate all vacua, the fermionic zero modes do not always annihilate the vaccum.  Instead, one finds:
\be
d^{(i)f,+ A}_0 \rpket^{(i)} ~=~ d^{(i)f,- A}_0 \rmket^{(i)} ~=~ 0
\ee
\be
d^{(i)f,+ A}_0 \rmket^{(i)}~\neq~ 0, ~~~\, d^{(i)f,- A}_0 \rpket^{(i)} ~\neq~ 0,
\ee
where the copy index $(i)$ is not summed over.  For more details about the behavior of these zero modes, see Appendix \ref{RVN}.

One can of course construct modes which live before the two twist insertions, as well as modes which live between the two twists.  Since we begin with the vacuum state as our in state before the twists and we do not need the state between the two twists in an explicit way, we will not write the modes in these regions.   We do however require modes natural to the NS vacuum in the $t$ plane.  These are
\bea
\tilde{\a}_{A\dot A,n} &=& {1\over 2\pi} \oint\limits_{t=0} \partial_t X_{A\dot A}(t) t^n \diff t \label{BosontMode}\\
\tilde{d}^{\a A}_r &=& {1\over 2\pi i} \oint\limits_{t=0} \psi^{\a A}(t) t^{r-\h} \diff t, \label{FermiontMode}
\eea
with commutation relations
\begin{eqnarray}
\left[\tilde{\a}_{A\dot{A},m},\tilde{\a}_{A\dot{A},n}\right]=-\e_{AB}\e_{\dot{A}\dot{B}}m\delta_{n+m,0} \label{bosoncommutation}\\
\left\{\tilde{d}^{\a A}_m,\tilde{d}^{\a A}_n\right\}=-\e^{\a\b}\e^{AB}\delta_{n+m,0}\;. \label{fermioncommutation}
\end{eqnarray}

As a final remark, there is a subtle concern that needs to be addressed when we wish to map the cylinder modes into the $t$ plane.  Since we have chosen the final copy 1 to map to large $t$, copy 1 modes must always come to the left of copy 2 modes in the $t$ plane.  As such, when we apply annihilation modes to probe the deformed state we will always choose to place all copy 1 modes to the left of all copy 2 modes.  This has no effect on the bosonic calculation because bosonic annihilators commute, but it is important to account for this convention when working with fermions.

\subsection{The general form}  
Looking at the SU(2) charges of the twist operators and the initial vacuum state, it is clear that $|\chi(w_1,w_2)\rangle$ is overall neutral.   One may also expect the final state to contain only pairs of excitations in the form of an exponential similar to the one-twist result of \cite{acm1}.  However, this cursory analysis seems to indicate that part of $|\chi(w_1,w_2)\rangle$ could live in the $NS$ sector, as the vacuum $\nsnsket$ is neutral.  We will perform the computations for both R and NS sectors in the final state, but argue at the end that only the R sector amplitudes are relevant for the physical system under study.

  In \cite{acm1} it was noted that the state produced after the twist had an exponential form, and a general argument was given for why this must be the case for the bosons of the theory.  This proof extends trivially to the use of arbitrarily many twists.  In Appendix \ref{GeneralFormAppendix} we present a similar proof for the fermionic contribution, which again extends to arbitrarily many twists and also applies to both the NS and Ramond sectors.  Keeping both sectors in the final state for the moment, we  note that $|\chi(w_1,w_2)\rangle$ can be written as:
\bea \label{generalform}
|\chi(w_1,w_2)\rangle & \!= \!& C_{NS}(w_1,w_2)\text{exp}\left [ \sum_{(i),(j)}\sum_{k,l > 0}\g^{B(i)(j)}_{kl} \left ( -\a^{(i)f}_{++,-k}\a^{(j)f}_{--,-l} + \a^{(i)f}_{+-,-k}\a^{(j)f}_{-+,-l} \right ) \right ] \nn
&&\!\!\!\!\!\!\!\!{}\times \text{exp}\left [ \sum_{(i),(j)}\sum_{r,s > 0}\g^{F(i)(j)}_{NS,rs} \left ( d^{(i)f,++}_{-r}d^{(j)f,--}_{-s} - d^{(i)f,+-}_{-r}d^{(j)f,-+}_{-s} \right ) \right ] \nsnsket \nn
&&\!\!\!\!\!\!\!\!{}+ C_{R+-}(w_1,w_2)\text{exp}\left [ \sum_{(i),(j)}\sum_{k,l > 0}\g^{B(i)(j)}_{kl} \left ( -\a^{(i)f}_{++,-k}\a^{(j)f}_{--,-l} + \a^{(i)f}_{+-,-k}\a^{(j)f}_{-+,-l} \right ) \right ] \nn
&&\!\!\!\!\!\!\!\!{}\times \text{exp}\left [ \sum_{(i),(j)}\sum_{k,l \geq 0}\g^{F(i)(j)}_{R+-,kl} \left ( d^{(i)f,++}_{-k}d^{(j)f,--}_{-l} - d^{(i)f,+-}_{-k}d^{(j)f,-+}_{-l} \right ) \right ] \rpmket, \nn
\label{chi state}
\eea
where the sums over the mode indices for the fermions in the Ramond sector include zero modes only when those modes do not annihilate the vacuum $\rpmket$.  Note that the various $\g$ coefficients also depend on the twist insertion points $w_1$ and $w_2$.

From here it is fairly straightforward to apply particular capping states with only a single pair of modes so as to pick out each specific $\g$ coefficient from $|\chi(w_1,w_2)\rangle$.  Taking a ratio of amplitudes, we can isolate these coefficients in a manner that is independent of other factors, such as the overall coefficients $C$.  We thus find:
\bea
\g^{B(i)(1)}_{mn} & = & -{1\over mn} {\nsnsbra \a^{(1)f}_{++,n}\a^{(i)f}_{--,m}|\chi(w_1,w_2)\rangle \over \nsnsbra \chi(w_1,w_2)\rangle} \label{BosonCylinderRatio1} \\
\g^{B(i)(2)}_{mn} & = & -{1\over mn} {\nsnsbra \a^{(i)f}_{--,m}\a^{(2)f}_{++,n}|\chi(w_1,w_2)\rangle \over \nsnsbra \chi(w_1,w_2)\rangle} \label{BosonCylinderRatio2} \\
\g^{F(i)(1)}_{NS,rs} & = & {\nsnsbra d^{(1)f,++}_{r}d^{(i)f,--}_{s}|\chi(w_1,w_2)\rangle \over \nsnsbra \chi(w_1,w_2)\rangle} \label{FermionNSCylinderRatio1} \\
\g^{F(i)(2)}_{NS,rs} & = & -{\nsnsbra d^{(i)f,--}_{s}d^{(2)f,++}_{r}|\chi(w_1,w_2)\rangle \over \nsnsbra \chi(w_1,w_2)\rangle} \label{FermionNSCylinderRatio2} \\
\g^{F(i)(1)}_{R+-,mn} & = & {\rpmbra d^{(1)f,++}_{n}d^{(i)f,--}_{m}|\chi(w_1,w_2)\rangle \over \rpmbra \chi(w_1,w_2)\rangle}\label{FermionRCylinderRatio1} \\
\g^{F(i)(2)}_{R+-,mn} & = & -{\rpmbra d^{(i)f,--}_{m}d^{(2)f,++}_{n}|\chi(w_1,w_2)\rangle \over \rpmbra \chi(w_1,w_2)\rangle},\label{FermionRCylinderRatio2}
\eea
where the minus signs in the fourth and sixth lines come from the fact that the fermion annihilators anticommute.  Note that the bosonic coefficients do not depend on the vacuum the state is built upon.  This is because the bosonic modes are unaffected by spectral flow, so they behave identically in both sectors.  Indeed, one can calculate these same bosonic coefficients by capping with a Ramond state instead.

From these expressions, we must perform the series of coordinate maps outlined in Section \ref{CoordinateMaps}.  This procedure removes all of the multivalued regions, though the Ramond vacua and twist insertions still bring spin fields.  We deal with these spin fields in the next subsection.

\subsection{Spectral flows}
We have two sectors we wish to cap with, the NS sector and the R sector.  The behavior of the bosonic modes is independent of the sector they act on, and as such we do not have to worry about which spectral flows we perform when it comes to the bosonic coefficients.  Thus we need only apply the coordinate maps to the right side of (\ref{BosonCylinderRatio1}) and (\ref{BosonCylinderRatio2}).  Doing so yields:
\bea
\g^{B(i)(1)}_{mn} & = & -{1\over mn} {\nstbra \a'^{(1)f}_{++,n}\a'^{(i)f}_{--,m}\nstket \over \nstbra \nstclose} \label{BosonCoverRatio1}\\
\g^{B(i)(2)}_{mn} & = & -{1\over mn} {\nstbra \a'^{(i)f}_{--,m}\a'^{(2)f}_{++,n}\nstket \over \nstbra \nstclose}, \label{BosonCoverRatio2}
\eea
where the primes denote that the modes have been altered by the coordinate shifts.

\begin{figure}[tbh]
\begin{center}
\includegraphics[width=0.4\columnwidth]{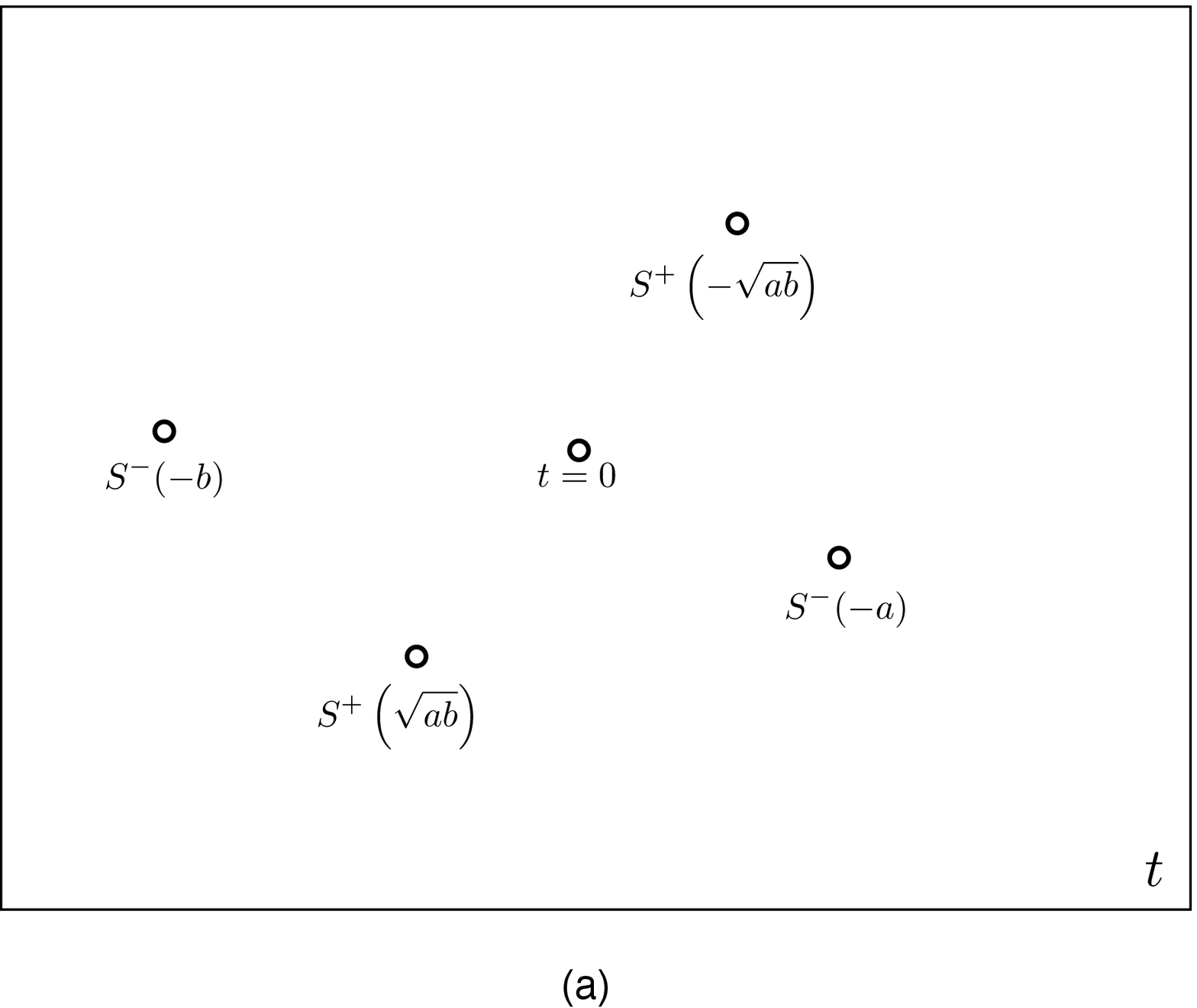} $\qquad\qquad$ \includegraphics[width=0.4\columnwidth]{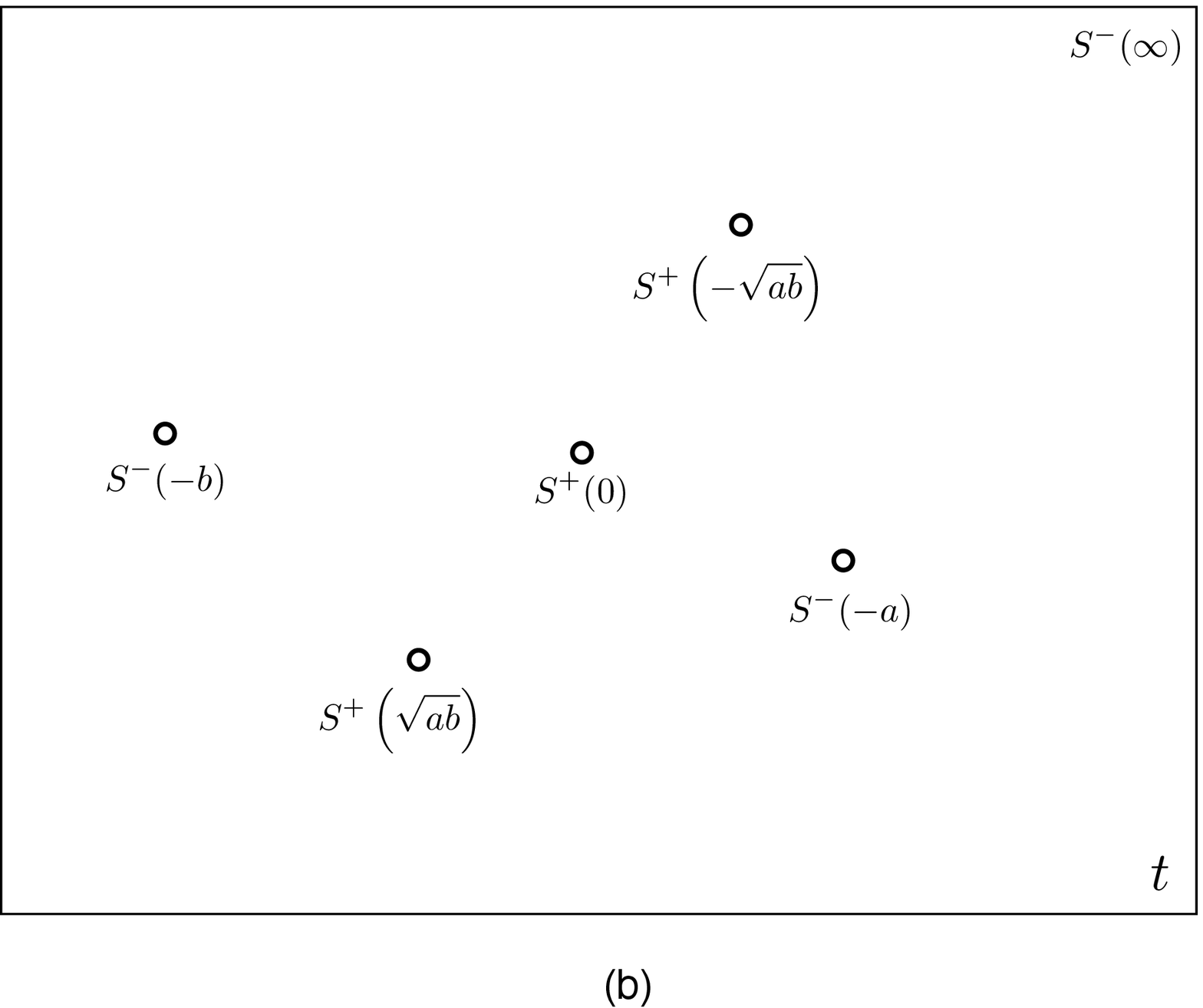}
\end{center}
\caption{Here we illustrate the insertions of all the spin fields in the $t$ plane when capping with the NS (a) and Ramond (b) sectors.}
\label{SpinInsertionFigure}
\end{figure}

For the fermionic coefficients, it is important to specify which sector we are building upon because the Ramond sector brings in additional spin fields.  Both cases are illustrated in Figure \ref{SpinInsertionFigure}.

When building upon the NS sector, the $t$-plane contains the following spin fields:
\bea\label{NSSpinLocations}
S^-(-a) &\text{from}& \rmket^{(1)} \nn
S^-(-b) &\text{from}& \rmket^{(2)} \nn
S^+(-\sqrt{ab}) &\text{from}& \s_2^+(w_1) \nn
S^+(\sqrt{ab}) &\text{from}& \s_2^+(w_2).
\eea
To remove these spin fields we must perform spectral flows at each of these points.  We spectral flow by $\a=+1$ at the locations of the $S^-$ spin fields, and by $\a=-1$ at the locations of the $S^+$ spin fields.  Since this combination contains two spectral flows in each direction all at finite points, there is no net effect at infinity.  This is important as we do not have any insertions at infinity.  After these spectral flows we can close all punctures in the $t$ plane, allowing us to smoothly deform the contours of the fermion modes.  Recalling (\ref{FermionNSCylinderRatio1}) and (\ref{FermionNSCylinderRatio2}), we find:
\bea
\g^{F(i)(1)}_{NS,rs} & = & {\nstbra d'^{(1)f,++}_{s}d'^{(i)f,--}_{r}\nstket \over \nstbra \nstclose} \label{FermionNSCoverRatio1}\\
\g^{F(i)(2)}_{NS,rs} & = & -{\nstbra d'^{(i)f,--}_{r}d'^{(2)f,++}_{s}\nstket \over \nstbra \nstclose}, \label{FermionNSCoverRatio2}
\eea
where here the primes denote that the modes have been altered by both the coordinate shifts and the four spectral flows given in (\ref{NSSpinLocations}).

When building upon the Ramond sector, we also have spin fields from the capping state $\rpmbra$.  Recall that for out states, copy 1 maps to $t=\infty$ while copy $2$ maps to the origin.  Also note that the lower-index charge indicated in the bra denotes that it is the conjugate of the ket of that charge, and thus has the opposite charge.  We now see that in addition to the spin fields encountered when building upon the NS vacuum, we also have:
\bea\label{RSpinLocations}
S^-(\infty) &\text{from}& {}^{(1)}\rpbra \nn
S^+(0) &\text{from}& {}^{(2)}\rmbra.
\eea
Both of these spin fields can be remove with a single spectral flow by $\a=-1$ units at the origin of the $t$ plane.  Recalling (\ref{FermionRCylinderRatio1}) and (\ref{FermionRCylinderRatio2}), we find:
\bea
\g^{F(i)(1)}_{R+-,mn} & = & {\nstbra \hat{d}^{(1)f,++}_{n}\hat{d}^{(i)f,--}_{m}\nstket \over \nstbra\nstclose}\label{FermionRCoverRatio1}\\
\g^{F(i)(2)}_{R+-,mn} & = & -{\nstbra \hat{d}^{(i)f,--}_{m}\hat{d}^{(2)f,++}_{n}\nstket \over \nstbra\nstclose},\label{FermionRCoverRatio2}
\eea
where the hats indicate that we have performed all of the coordinate maps and spectral flows required by (\ref{NSSpinLocations}) along with an additional spectral flow by $\a=-1$ units at the origin of the $t$ plane.

We now have useful expressions for the various $\g$ coefficients.  By applying the required coordinate shifts and spectral flows to the bosonic and fermionic modes and expanding the results in terms of modes natural to the $t$ plane, we can use the known commutation relations to evaluate equations (\ref{BosonCoverRatio1}) through (\ref{FermionRCoverRatio2}).  We perform these calculations in the following sections.

\section{Computation of $\g^{B(i)(j)}_{mn}$}\label{Boson}

Here we calculate the bosonic $\g$ coefficients using the relations (\ref{BosonCoverRatio1}) and (\ref{BosonCoverRatio2}).  To use these expressions, we first need to compute the $\a'$ modes in the $t$ plane.  Since the bosonic modes are unaffected by spectral flow, we need only apply the coordinate maps to the original modes on the cylinder.  We thus organize this section into three parts.  In the first part, we will apply the coordinate maps outlined in Section \ref{CoordinateMaps} to determine the expressions for the $\a'$ modes.  In the second part, we will expand the $\a'$ modes in terms of the modes natural to the $t$ plane.  In the final part, we will use these expressions in (\ref{BosonCoverRatio1}) and (\ref{BosonCoverRatio2}) to compute the bosonic $\g$ coefficients.

\subsection{Mapping the boson modes to the $t$ plane}
We wish to apply the various coordinate maps presented in Section \ref{CoordinateMaps} to the post-twist modes on the cylinder given in (\ref{BosonCylinderMode}).  Under a general coordinate transformation $w \to w'$, the bosonic field $X$ tranforms as:
\bea
X^{(i)}_{A\dot A}(w) &\to& {\diff w' \over \diff w} X^{(i)}_{A\dot A}\left ( w' \right ).
\eea
However, the jacobian brings a factor:
\bea\label{Jacobian}
\diff w &\to& \left ({\diff w' \over \diff w} \right )^{-1} \diff w'.
\eea
Thus the combination of the two behaves as:
\bea
X^{(i)}_{A\dot A}(w)\diff w &\to& X^{(i)}_{A\dot A}(w') \diff w'.
\eea

We first map the cylinder with coordinate $w$ to the plane with coordinate $z$ via
\be
z=e^w.
\ee
Since we do not make use of the initial or intermediate modes in this paper, we just write the modes after the second twist insertion ($|z|>e^{\tau_2}$).  Here we have a contour circling ${z=\infty}$:
\be
\alpha_{A\dot A, n}^{(1)f}\to{1\over 2\pi} \oint_{z=\infty} X_{A\dot A}^{(1)f}(z) z^{n} \diff z
\ee
\be
\alpha_{A\dot A, n}^{(2)f}\to{1\over 2\pi} \oint_{z=\infty} X_{A\dot A}^{(2)f}(z) z^{n} \diff z . 
\ee

We now proceed to the covering space $t$, where $X_{A\dot A}$ will be single-valued. Using the map defined earlier,
\be
z=\frac{(t+a)(t+b)}{t}, \label{cover}
\ee
and again only considering the modes after the second twist insertion, we have
\be
\alpha^{(1)f}_{A\dot A, n}\to \a'^{(1)f}_{A\dot A ,n}~=~{1\over 2\pi} \oint_{t=\infty}X_{A\dot A}(t) \left(\frac{(t+a)(t+b)}{t}\right)^{n} \diff t
\label{a1f}
\ee
\be
\alpha^{(2)f}_{A\dot A, n}\to \a'^{(2)f}_{A\dot A ,n}~=~-{1\over 2\pi} \oint_{t=0}X_{A\dot A}(t) \left(\frac{(t+a)(t+b)}{t}\right)^{n} \diff t.
\label{a2f}
\ee
Note the minus sign in (\ref{a2f}).  As mentioned earlier, the copy 2 values at $z=\infty$ map to $t=0$.  This map has the leading-order behavior $z \sim \tfrac{ab}{t}$, which means $arg(z) \sim -arg(t)$.  Thus the contour changes direction when we map to the origin of the $t$ plane.  This is the source of the minus sign.

\subsection{Expanding the Bosonic Modes}
In order to expand our $\a'$ modes in terms of $\tilde{\a}$, it suffices to expand the integrand of the former in powers of $t$.  This expansion must be performed in the region where the mode lives, $t\sim\infty$ for the copy 1 modes and $t\sim0$ for the copy 2 modes.  We perform the relevant expansions for each copy below.
\subsubsection*{Copy 1, \textbf{$t\sim\infty$}}
\bea
\left(\frac{(t+a)(t+b)}{t}\right)^{n}&=&t^{n}\left(1+at^{-1}\right)^{n}\left(1+bt^{-1}\right)^{n}\nn
&=&t^{n}\sum_{j,j'\geq 0}{}^{n}C_{j}{}^{n}C_{j'}a^{j}b^{j'}t^{-j-j'}\nn
&=&\sum_{j,j'\geq 0}{}^{n}C_{j}{}^{n}C_{j'}a^{j}b^{j'}t^{n-j-j'}.
\eea
\subsubsection*{Copy 2, \textbf{$t\sim0$}}
\bea
\left(\frac{(t+a)(t+b)}{t}\right)^{n}&=&t^{-n}a^{n}b^{n}\left(1+ta^{-1}\right)^{n}\left(1+tb^{-1}\right)^{n}\nn
&=&t^{-n}a^{n}b^{n}\sum_{j,j'\geq 0}{}^{n}C_{j}{}^{n}C_{j'}a^{-j}b^{-j'}t^{j+j'}\nn
&=&\sum_{j,j'\geq 0}{}^{n}C_{j}{}^{n}C_{j'}a^{n-j}b^{n-j'}t^{j+j'-n}.
\eea
Thus (\ref{a1f}) and (\ref{a2f}) become 
\bea
\a'^{(1),f}_{A\dot{A},n}&=&\sum_{j,j'\geq 0} {}^{n}C_{j}{}^{n}C_{j'}a^{j}b^{j'} {1\over 2\pi}\oint_{t=\infty}X_{A\dot A}(t)t^{n-j-j'}\diff t\nn
&=&\sum_{j,j'\geq 0} {}^{n}C_{j}{}^{n}C_{j'}a^{j}b^{j'} \tilde{\a}_{A\dot{A},n-j-j'} \label{AlphaOneExpansion}\\
\a'^{(2),f}_{A\dot{A},n}&=&-\sum_{j,j'\geq 0} {}^{n}C_{j}{}^{n}C_{j'}a^{n-j}b^{n-j'}{1\over 2\pi}\oint_{t=\infty}X_{A\dot A}(t)t^{j+j'-n}\diff t \nn
&=&-\sum_{j,j'\geq 0} {}^{n}C_{j}{}^{n}C_{j'}a^{n-j}b^{n-j'}\tilde{\a}_{A\dot{A},j+j'-n}. \label{AlphaTwoExpansion}
\eea

\subsection{Computing $\g^{B(i)(j)}_{mn}$}
In this section we compute the $\g^{B(i)(j)}_{mn}$ coefficients. From (\ref{BosonCylinderRatio1}) and (\ref{AlphaOneExpansion}) we find: 
\bea
\g^{B(1)(1)}_{mn}=-\frac{1}{mn}\sum_{j,j',k,k'\geq 0}{}^{n}C_{j}{}^{n}C_{j'}{}^{m}C_{k}{}^{m}C_{k'}a^{j+k}b^{j'+k'}\frac{\nstbra\tilde{\a}_{++,n-j-j'}\tilde{\a}_{--,m-k-k'}\nstket}{\nstbra\nstclose}.\nn\label{Gamma11TripleSum}
\eea
It is clear that the summand vanishes except when
\bea
m+n-j-j'-k-k'=0~&\implies& ~k'=m+n-j-j'-k \nn
n-j-j'> 0~&\implies&~n-j > j'. \label{constraints1}
\eea
Furthermore,
\bea
j'\geq 0~&\implies& n > j\nn
k'\geq 0 ~&\implies& ~ m+n-j-j'\geq k. \label{constraints2}
\eea
We can thus rewrite (\ref{Gamma11TripleSum}) as
\bea
\g^{B(1)(1)}_{mn}=\frac{1}{mn}\sum_{j=0}^{n-1}\sum_{j'=0}^{n-j-1}\sum_{k=0}^{m+n-j-j'}(n-j-j'){}^{n}C_{j}{}^{n}C_{j'}{}^{m}C_{k}{}^{m}C_{m+n-j-j'-k}a^{j+k}b^{m+n-j-k}.\nn
\eea
Following these same steps for the other cases, we find:
\bea
\g^{B(1)(2)}_{mn}=-\frac{1}{mn}\sum_{j=0}^{n-1}\sum_{j'=0}^{n-j-1}\sum_{k=0}^{m-n+j+j'}(n-j-j'){}^{n}C_{j}{}^{n}C_{j'}{}^{m}C_{k}{}^{m}C_{m-n+j+j'-k}a^{n-j+k}b^{m+j-k}\nn
\label{gamma12boson}
\eea
\bea
\g^{B(2)(1)}_{mn}=-\frac{1}{mn}\sum_{j=0}^{n-1}\sum_{j'=0}^{n-j-1}\sum_{k=0}^{m-n+j+j'}(n-j-j'){}^{n}C_{j}{}^{n}C_{j'}{}^{m}C_{k}{}^{m}C_{m-n+j+j'-k}a^{m+j-k}b^{n-j+k}\nn
\label{gamma21boson}
\eea
\bea
\g^{B(2)(2)}_{mn}=\frac{1}{mn}\sum_{j=0}^{n-1}\sum_{j'=0}^{n-j-1}\sum_{k=0}^{m+n-j-j'}(n-j-j'){}^{n}C_{k}{}^{n}C_{m+n-j-j'-k}{}^{m}C_{j}{}^{m}C_{j'}a^{m+n-j-k}b^{j+k}.\nn
\label{gamma22boson}
\eea
The physics here is symmetric under the interchange $a\leftrightarrow b$. One can explicitly check that this symmetry holds in the above expressions, though this fact is not immediately manifest. Using this symmetry, we find
\bea
\g^{B(2)(2)}_{mn}=\frac{1}{mn}\sum_{j=0}^{n-1}\sum_{j'=0}^{n-j-1}\sum_{k=0}^{m+n-j-j'}(n-j-j'){}^{n}C_{j}{}^{n}C_{j'}{}^{m}C_{k}{}^{m}C_{m+n-j-j'-k}a^{j+k}b^{m+n-j-k},\quad
\eea
and therefore
\bea
\g^{B(2)(2)}_{mn}=\g^{B(1)(1)}_{mn}.\label{BosonRelation1}
\eea
Similarly,
\bea
\g^{B(2)(1)}_{mn}=\g^{B(1)(2)}_{mn}.\label{BosonRelation2}
\eea

We will now bring the $\g^{B(i)(j)}_{mn}$ coefficients into a form that involves only one summation over a pair of hypergeometric functions, binomial coefficients, and $a$ and $b$ terms raised to the appropriate powers. We achieve this through a re-definition of the summation indices.  We present the computation for $\g^{B(1)(1)}_{mn}$ and simply skip to the results for the remaining cases.

For $\g^{B(1)(1)}_{mn}$, we begin with:
\bea
\g^{B(1)(1)}_{mn}=\frac{1}{mn}\sum_{j=0}^{n-1}\sum_{j'=0}^{n-j-1}\sum_{k=0}^{m+n-j-j'}(n-j-j'){}^{n}C_{j}{}^{n}C_{j'}{}^{m}C_{k}{}^{m}C_{m+n-j-j'-k}a^{j+k}b^{m+n-j-k}.\nn
\label{gamma11}
\eea
We now define a new summation index:
\bea
l=n-j-j'~&\implies&~j'=n-j-l,
\label{indexredefition}
\eea
where (\ref{constraints1}) and (\ref{constraints2}) now give
\bea
n-j-j' > 0 ~&\implies&~ l > 0 \nn
j'\geq 0 ~&\implies&~ n-l \geq j \nn
j,j'\geq 0 ~&\implies&~ n \geq l.
\eea
Applying these constraints, we find:
\bea
\g^{B(1)(1)}_{mn}=\frac{1}{mn}\sum_{l=1}^{n}lb^{m+n}\left(\sum_{j=0}^{n-l}{}^{n}C_{j}{}^{n}C_{n-j-l}a^{j}b^{-j}\right)\left(\sum_{k=0}^{m+l}{}^{m}C_{k}{}^{m}C_{m+l-k}a^{k}b^{-k}\right).
\label{Gamma11_2}
\eea
The $k$ and $j$ sums can be evaluated in \textit{Mathematica}\footnote{Mathematica sometimes has trouble dealing with the fact that some of the binomial coefficients vanish when the second argument is negative.  It is occasionally necessary to alter the range of the $j$ sum to explicitly remove these vanishing terms in order to obtain the full result.}, resulting in:\footnote{While it is not immediately apparent, these expressions are symmetric under $a \leftrightarrow b$, as expected.}
\bea
\g^{B(1)(1)}_{mn}&=&\frac{1}{mn}\sum_{l=1}^{n}l{}^{n}C_{l}{}^{m}C_{l}a^{l}b^{m+n-l}{}_2F_1\left(-n,l-n;l+1;\frac{a}{b}\right){}_2F_1\left(-m,l-m;l+1;\frac{a}{b}\right).\nn
\eea

The other $\g^{B(i)(j)}$ coefficients are calculated in the same manner, using the same index re-definition.  We present the results here.
\bea
\g^{B(1)(2)}_{mn}&=&-\frac{1}{mn}\sum_{l=1}^{n}l{}^{m}C_{l}{}^{n}C_{l}a^{n}b^{m}{}_2F_1\left(-m,l-m;l+1;\frac{a}{b}\right) {}_2F_1\left(-n,l-n;l+1;\frac{b}{a}\right)\nn\nn
\g^{B(2)(1)}_{mn}&=&-\frac{1}{mn}\sum_{l=1}^{n}l{}^{m}C_{l}{}^{n}C_{l}a^{m}b^{n}{}_2F_1\left(-m,l-m;l+1;\frac{b}{a}\right){}_2F_1\left(-n,l-n;l+1;\frac{a}{b}\right)\nn\nn
\g^{B(2)(2)}_{mn}&=&\frac{1}{mn}\sum_{l=1}^{n}l{}^{n}C_{l}{}^{m}C_{l} a^{m+n-l}b^{l}{}_2F_1\left(-m,l-m;l+1;\frac{b}{a}\right){}_2F_1\left(-n,l-n;l+1;\frac{b}{a}\right). \nn
\eea

In addition to (\ref{BosonRelation1}) and (\ref{BosonRelation2}), the above forms allow us to show the additional relation:
\bea
\g^{B(1)(1)}_{mn} &=& - \g^{B(1)(2)}_{mn}.
\eea
This relationship is shown explicitly in Appendix \ref{GammaRelations}.  We then have only a single linearly independent bosonic coefficient.

\section{Computation of $\g^{F(i)(j)}_{NS,mn}$}\label{sec:f}
Here we perform the computation of the non-physical $\g^{F(i)(j)}_{NS,mn}$ coefficients, where the $NS$ indicates that we are capping the squeezed state with the non-physical $NS$ vacuum for both copy $1$ and copy $2$. We proceed in a manner identical to the boson case, except that the fermion modes are affected by the spectral flows which were needed to remove the spin fields in (\ref{NSSpinLocations}).

\subsection{Mapping the fermion modes to the $t$ plane}
The NS sector fermion modes on the cylinder were given in (\ref{FermionNSCylinderMode}).  Here we map these modes to the $t$ plane by way of the $z$ plane.

Under a general coordinate transformation $w\to w'$, the fermion field $\psi$ transforms as:
\bea
\psi^{(i),\a A}(w) \to \left ( {\diff w' \over \diff w}\right )^{\h} \psi^{(i),\a A}(w'),
\eea
while the Jacobian again brings the factor described in (\ref{Jacobian}).  Thus the two together transform as:
\bea
\psi^{(i),\a A}(w)\diff w \to \left ( {\diff w' \over \diff w}\right )^{-\h} \psi^{(i),\a A}(w')\diff w'.
\eea
Thus when mapping to the $z$ plane via $z = e^w$, we have
\bea
\psi^{\alpha A}(w)\diff w \to z^{-1/2}\psi^{\alpha A}(z) \diff z,
\eea
and the fermionic modes become:
\bea
d_{r}^{(1)f,\a A}&\to&\frac{1}{2\pi i}\int_{\sigma=0}^{2\pi} 
\psi^{(1)f,\a A}(z)z^{r-1/2}\diff z\nn
d_{r}^{(2)f,\a A}&\to&\frac{1}{2\pi i}\int_{\sigma=0}^{2\pi}
\psi^{(2)f,\a A}(z)z^{r-1/2}\diff z.\nn
\eea

We now move to the covering $t$ plane in the same manner, using the map (\ref{tmap}).  We then find:  
\bea
d_{r}^{(1)f,\a A}&\to&\frac{1}{2\pi i}\int_{t=\infty}
\psi^{(1)f,\a A}(t)t^{-r-1/2}(t+\sqrt{ab})^{1/2}(t-\sqrt{ab})^{1/2}(t+a)^{r-1/2}(t+b)^{r-1/2}\diff t\nn
d_{r}^{(2)f,\a A}&\to&\!-\frac{1}{2\pi i}\int_{t=0}
\psi^{(2)f,\a A}(t)t^{-r-1/2}(t+\sqrt{ab})^{1/2}(t-\sqrt{ab})^{1/2}(t+a)^{r-1/2}(t+b)^{r-1/2}\diff t,\nn
\eea
where again the copy 2 mode gains a minus sign from the reversal of the direction of the contour as $z \sim t^{-1}$.

\subsection{Applying the spectral flows}
We now perform the spectral flows required to eliminate the spin fields in the $t$ plane (\ref{NSSpinLocations}).  Under a spectral flow by $\a$ units at a point $t_0$, the fermion field $\psi$ behaves as:
\bea
\psi^{\pm A}(t)\to(t-t_0)^{\mp \h \a}\psi^{\pm A}(t).
\eea
We perform each of the four spectral flows in turn.

\subsubsection*{\underline{$\a = 1$ at $t=-b$}}
\bea
d_{r}^{(1)f,+A}&\to&\frac{1}{2\pi i}\int_{t=\infty}\psi^{(1)f,+A}(t)t^{-r-1/2}(t+\sqrt{ab})^{1/2}(t-\sqrt{ab})^{1/2}(t+a)^{r-1/2}(t+b)^{r-1}\diff t\nn
d_{r}^{(1)f,-A}&\to&\frac{1}{2\pi i}\int_{t=\infty}\psi^{(1)f,-A}(t)t^{-r-1/2}(t+\sqrt{ab})^{1/2}(t-\sqrt{ab})^{1/2}(t+a)^{r-1/2}(t+b)^{r}\diff t\nn
d_{r}^{(2)f,+A}&\to&-\frac{1}{2\pi i}\int_{t=0}\psi^{(2)f,+A}(t)t^{-r-1/2}(t+\sqrt{ab})^{1/2}(t-\sqrt{ab})^{1/2}(t+a)^{r-1/2}(t+b)^{r-1}\diff t\nn
d_{r}^{(2)f,-A}&\to&-\frac{1}{2\pi i}\int_{t=0}\psi^{(2)f,-A}(t)t^{-r-1/2}(t+\sqrt{ab})^{1/2}(t-\sqrt{ab})^{1/2}(t+a)^{r-1/2}(t+b)^{r}\diff t.\nn
\eea

\subsubsection*{\underline{$\a = 1$ at $t=-a$}}
\bea
d_{r}^{(1)f,+A}&\to&\frac{1}{2\pi i}\int_{t=\infty}\psi^{(1)f,+A}(t)t^{-r-1/2}(t+\sqrt{ab})^{1/2}(t-\sqrt{ab})^{1/2}(t+a)^{r-1}(t+b)^{r-1}\diff t\nn
d_{r}^{(1)f,-A}&\to&\frac{1}{2\pi i}\int_{t=\infty}\psi^{(1)f,-A}(t)t^{-r-1/2}(t+\sqrt{ab})^{1/2}(t-\sqrt{ab})^{1/2}(t+a)^{r}(t+b)^{r}\diff t\nn
d_{r}^{(2)f,+A}&\to&-\frac{1}{2\pi i}\int_{t=0}\psi^{(2)f,+A}(t)t^{-r-1/2}(t+\sqrt{ab})^{1/2}(t-\sqrt{ab})^{1/2}(t+a)^{r-1}(t+b)^{r-1}\diff t\nn
d_{r}^{(2)f,-A}&\to&-\frac{1}{2\pi i}\int_{t=0}\psi^{(2)f,-A}(t)t^{-r-1/2}(t+\sqrt{ab})^{1/2}(t-\sqrt{ab})^{1/2}(t+a)^{r}(t+b)^{r}\diff t.\nn
\eea

\subsubsection*{\underline{$\a = -1$ at $t=-\sqrt{ab}$}}
\bea
d_{r}^{(1)f,+A}&\to&\frac{1}{2\pi i}\int_{t=\infty}\psi^{(1)f,+A}(t)t^{-r-1/2}(t+\sqrt{ab})(t-\sqrt{ab})^{1/2}(t+a)^{r-1}(t+b)^{r-1}\diff t\nn
d_{r}^{(1)f,-A}&\to&\frac{1}{2\pi i}\int_{t=\infty}\psi^{(1)f,-A}(t)t^{-r-1/2}(t-\sqrt{ab})^{1/2}(t+a)^{r}(t+b)^{r}\diff t\nn
d_{r}^{(2)f,+A}&\to&-\frac{1}{2\pi i}\int_{t=0}\psi^{(2)f,+A}(t)t^{-r-1/2}(t+\sqrt{ab})(t-\sqrt{ab})^{1/2}(t+a)^{r-1}(t+b)^{r-1}\diff t\nn
d_{r}^{(2)f,-A}&\to&-\frac{1}{2\pi i}\int_{t=0}\psi^{(2)f,-A}(t)t^{-r-1/2}(t-\sqrt{ab})^{1/2}(t+a)^{r}(t+b)^{r}\diff t.
\eea

\subsubsection*{\underline{$\a = -1$ at $t=\sqrt{ab}$}}
\bea
d_{r}^{(1)f,+A}&\to&d'^{(1)f,+A}_{r}~=~\frac{1}{2\pi i}\int_{t=\infty}\psi^{(1)f,+A}(t)t^{-r-1/2}(t^{2}-ab)(t+a)^{r-1}(t+b)^{r-1}\diff t\nn
d_{r}^{(1)f,-A}&\to&d'^{(1)f,-A}_{r}~=~\frac{1}{2\pi i}\int_{t=\infty}\psi^{(1)f,-A}(t)t^{-r-1/2}(t+a)^{r}(t+b)^{r}\diff t\nn
d_{r}^{(2)f,+A}&\to&d'^{(2)f,+A}_{r}~=~-\frac{1}{2\pi i}\int_{t=0}\psi^{(2)f,+A}(t)t^{-r-1/2}(t^{2}-ab)(t+a)^{r-1}(t+b)^{r-1}\diff t\nn
d_{r}^{(2)f,-A}&\to&d'^{(2)f,-A}_{r}~=~-\frac{1}{2\pi i}\int_{t=0}\psi^{(2)f,-A}(t)t^{-r-1/2}(t+a)^{r}(t+b)^{r}\diff t.\label{fermionmodes}
\eea
We have now removed all spin field insertions in the $t$-plane and can close all punctures with the state $\nstket$.  This allows us to smoothly deform our contours.

\subsection{Expanding the fermion modes}
The operators $d'^{(1)f,\pm A}_r$ are given by contour integrals at large $t$ while the operators $d'^{(2)f,\pm A}_r$ are given by contour integrals around $t=0$. We need to express these operators in terms of the fermion modes natural to the $t$ plane, which are contour integrals of powers of $t$.  We thus expand the integrand of each mode in powers of $t$ around its appropriate region in the $t$-plane.

\subsubsection*{$\underline{d'^{(1)f,+A}_{r}}$}
\bea
t^{-r-1/2}(t^{2}-ab)(t+a)^{r-1}(t+b)^{r-1} &=& t^{r-5/2}(t^{2}-ab)\sum_{q,q'\geq 0}^{\infty} {}^{r-1}C_q{}^{r-1}C_{q'} a^q b^{q'} t^{-q-q'} \nn
&=&\sum_{q,q'\geq 0} {}^{r-1}C_q{}^{r-1}C_{q'} a^q b^{q'} t^{r-q-q'-1/2}\nn
&&-\sum_{q,q'\geq 0} {}^{r-1}C_q{}^{r-1}C_{q'} a^{q+1} b^{q'+1} t^{r-q-q'-5/2}. \nn
\eea

\subsubsection*{$\underline{d'^{(1)f,-A}_{r}}$}
\bea
t^{-r-1/2}(t+a)^{r}(t+b)^{r} &=& t^{r-1/2}\sum_{q,q'\geq 0} {}^{r}C_q{}^{r}C_{q'} a^{q} b^{q'} t^{-q-q'} \nn
&=&\sum_{q,q\geq 0} {}^{r}C_q{}^{r}C_{q'} a^{q} b^{q'} t^{r-q-q'-1/2}.
\eea
\subsubsection*{$\underline{d'^{(2)f,+A}_{r}}$}
\bea
t^{-r-1/2}(t^{2}-ab)(t+a)^{r-1}(t+b)^{r-1} &=& t^{-r-1/2}(t^{2}-ab)a^{r-1}b^{r-1}\sum_{q,q'\geq 0} {}^{r-1}C_q{}^{r-1}C_{q'} a^{-q} b^{-q'} t^{q+q'} \nn
&=&\sum_{q,q'\geq 0}{}^{r-1}C_q{}^{r-1}C_{q'} a^{r-q-1} b^{r-q'-1} t^{q+q'-r+3/2}\nn
&&-\sum_{q,q'\geq 0} {}^{r-1}C_q{}^{r-1}C_{q'} a^{r-q} b^{r-q'} t^{q+q'-r-1/2}. \nn
\eea
\subsubsection*{$\underline{d'^{(2)f,-A}_{r}}$}
\bea
t^{-r-1/2}(t+a)^{r}(t+b)^{r} &=& t^{-r-1/2}a^r b^r\sum_{q,q'\geq 0} {}^{r}C_q{}^{r}C_{q'} a^{-q} b^{-q'} t^{q+q'} \nn
&=&\sum_{q,q'\geq 0} {}^{r}C_q{}^{r}C_{q'} a^{r-q} b^{r-q'} t^{q+q'-r-1/2}.
\eea

With these expansions, (\ref{fermionmodes}) becomes:
\bea
d'^{(1)f,+A}_{r}&=&\sum_{q,q'\geq 0} {}^{r-1}C_q{}^{r-1}C_{q'} a^q b^{q'} \frac{1}{2\pi i}\int_{t=\infty}\psi^{(1)f,+A}(t)t^{r-q-q'-1/2}\diff t\nn 
&&-\sum_{q,q\geq 0} {}^{r-1}C_q{}^{r-1}C_{q'} a^{q+1} b^{q'+1}\frac{1}{2\pi i}\int_{t=\infty}\psi^{(1)f,+A}(t)t^{r-q-q'-5/2}\diff t\nn
d'^{(1)f,-A}_{r}&\to&\sum_{q,q'\geq 0} {}^{r}C_q{}^{r}C_{q'} a^{q} b^{q'}\frac{1}{2\pi i}\int_{t=\infty}\psi^{(1)f,-A}(t)t^{r-q-q'-1/2}\diff t\nn
d'^{(2)f,+A}_{r}&\to&-\sum_{q,q'\geq 0}{}^{r-1}C_q{}^{r-1}C_{q'} a^{r-q-1} b^{r-q'-1} \frac{1}{2\pi i}\int_{t=0}\psi^{(2)f,+A}(t)t^{q+q'-r+3/2}\diff t\nn
&&+\sum_{q,q'\geq 0} {}^{r-1}C_q{}^{r-1}C_{q'} a^{r-q} b^{r-q'}\frac{1}{2\pi i}\int_{t=0}\psi^{(2)f,+A}(t)t^{q+q'-r-1/2}\diff t\nn
d'^{(2)f,-A}_{r}&\to&-\sum_{q,q'\geq 0} {}^{r}C_q{}^{r}C_{q'} a^{r-q} b^{r-q'}\frac{1}{2\pi i}\int_{t=0}\psi^{(2)f,-A}(t)t^{q+q'-r-1/2}\diff t.
\eea
In terms of modes natural to the $t$-plane, we have:
\bea
d'^{(1)f,+A}_{r}&=&\sum_{q,q'\geq 0} {}^{r-1}C_q{}^{r-1}C_{q'} a^q b^{q'} \tilde{d}^{+A}_{r-q-q'}-\sum_{q,q'\geq 0} {}^{r-1}C_q{}^{r-1}C_{q'} a^{q+1} b^{q'+1}\tilde{d}_{r-q-q'-2}^{+A}\nn
d'^{(1)f,-A}_{r}&=&\sum_{q,q'\geq 0} {}^{r}C_q{}^{r}C_{q'} a^{q} b^{q'}\tilde{d}^{-A}_{r-q-q'}\nn
d'^{(2)f,+A}_{r}&=&-\sum_{q,q'\geq 0}{}^{r-1}C_q{}^{r-1}C_{q'} a^{r-q-1} b^{r-q'-1} \tilde{d}^{+A}_{q+q'-r+2}+\sum_{q,q'\geq 0} {}^{r-1}C_q{}^{r-1}C_{q'} a^{r-q} b^{r-q'}\tilde{d}^{+A}_{q+q'-r}\nn
d'^{(2)f,-A}_{r}&=&-\sum_{q,q'\geq 0} {}^{r}C_q{}^{r}C_{q'} a^{r-q} b^{r-q'}\tilde{d}^{-A}_{q+q'-r}.\label{tplanemodes}
\eea

\subsection{Computing $\g^{F(i)(j)}_{NS,rs}$}
Here we shall compute the $\g^{F(i)(j)}_{NS}$ coefficients. We present the computation for $\g^{F(1)(1)}_{NS}$ explicitly. The other computations are performed in the same way, and we give only the results. Using (\ref{FermionNSCoverRatio1}) and (\ref{tplanemodes}), we find:
\bea
\g^{F(1)(1)}_{NS,rs}&=&\sum_{p,p',q,q'\geq 0}{}^{s-1}C_q{}^{s-1}C_{q'}{}^{r}C_p{}^{r}C_{p'}a^{q+p}b^{q'+p'}~\frac{\nstbra\tilde{d}^{++}_{s-q-q'}\tilde{d}^{--}_{r-p-p'}\nstket}{\nstbra\nstclose_{t}}\nn
&&{}-\sum_{p,p',q,q'\geq 0}{}^{s-1}C_q{}^{s-1}C_{q'}{}^{r}C_p{}^{r}C_{p'}a^{q+p+1}b^{q'+p'+1}~\frac{\nstbra\tilde{d}_{s-q-q'-2}^{++}\tilde{d}^{--}_{r-p-p'}\nstket}{\nstbra\nstclose_{t}}.\nn
\label{gamma11NS}
\eea
For the first term, it is clear that the summand vanishes except when:
\bea
r+s-p-p'-q-q'=0~&\implies& ~p'=r+s-q-q'-p\nn
s-q-q' > 0 ~&\implies& ~ s-q > q'.
\eea
Furthermore,
\bea
q'\geq 0 ~&\implies&~ s > q \nn
p' \geq 0 ~&\implies&~ r+s-q-q' \geq p.
\eea
We can thus rewrite (\ref{gamma11NS}) as:
\bea
\g^{F(1)(1)}_{NS,rs}&=&-\sum_{q=0}^{\lfloor s\rfloor}\sum_{q'=0}^{\lfloor s\rfloor-q}\,\sum_{p=0}^{r+s-q-p}{}^{s-1}C_q{}^{s-1}C_{q'}{}^{r}C_p{}^{r}C_{r+s-q-q'-p}a^{q+p}b^{r+s-q-q'} \nn
&&{}-\sum_{p,p',q,q'\geq 0}{}^{s-1}C_q{}^{s-1}C_{q'}{}^{r}C_p{}^{r}C_{p'}a^{q+p+1}b^{q'+p'+1}~\frac{\langle 0_{NS}|\tilde{d}_{s-q-q'-2}^{++}\tilde{d}^{--}_{r-p-p'}|0_{NS}\rangle_{t}}{\nstbra\nstclose}.\nn
\label{gamma11NS_2}
\eea
Applying the same line of reasoning for the second term, we find:
\bea
r+s-p-p'-q-q'-2=0~&\implies& ~p'=r+s-q-q'-p-2\nn
s-q-q'-2 >0~&\implies& s-q-2> q'\nn
p'\geq 0 ~&\implies& ~ r+s-q-q'-2\geq p\nn
q'\geq 0~&\implies& s-2> q.
\eea
So (\ref{gamma11NS_2}) becomes
\bea
\g^{F(1)(1)}_{NS,rs}&=&-\sum_{q=0}^{\lfloor s\rfloor}\sum_{q'=0}^{\lfloor s\rfloor-q}\,\sum_{p=0}^{r+s-q-p}{}^{s-1}C_q{}^{s-1}C_{q'}{}^{r}C_p{}^{r}C_{r+s-q-q'-p}a^{q+p}b^{r+s-q-q'}\nn
&&+\sum_{q=0}^{\lfloor s\rfloor-2}\sum_{q'=0}^{\lfloor s\rfloor-q-2}\,\sum_{p=0}^{r+s-q-q'-2}{}^{s-1}C_q{}^{s-1}C_{q'}{}^{r}C_p{}^{r}C_{r+s-q-q'-p-2}a^{q+p+1}b^{r+s-p-q-1},\nn
\eea
where the second term is instead zero when the sums' ranges are invalid ($s \leq \tfrac{3}{2}$).  We shall adopt this convention generally.  Henceforth, all sums are to be treated as zero when the given range is invalid. 

The other cases proceed along the same lines.
\bea
\g^{F(1)(2)}_{NS,rs}&=&-\sum_{q=0}^{\lfloor s\rfloor-2}\sum_{q'=0}^{\lfloor s\rfloor-q-2}\,\sum_{p=0}^{r-s+q+q'+2}{}^{s-1}C_{q}{}^{s-1}C_{q'}{}^{r}C_{p}{}^{r}C_{r-s+q+q'-p+2}a^{s-q+p-1}b^{r+q-p+1}\nn
&&+\sum_{q=0}^{\lfloor s\rfloor}\sum_{q'=0}^{\lfloor s\rfloor-q}\,\sum_{p=0}^{r-s+q+q'}{}^{s-1}C_{q}{}^{s-1}C_{q'}{}^{r}C_{p}{}^{r}C_{r-s+q+q'-p}a^{s-q+p}b^{r+q-p}
\label{gamma12NS}
\eea
\bea
\g^{F(2)(1)}_{NS,rs}&=&-\sum_{q=0}^{\lfloor s\rfloor-2}\sum_{q'=0}^{\lfloor s\rfloor-q-2}\,\sum_{p=0}^{r-s+q+q'+2}{}^{s-1}C_{q}{}^{s-1}C_{q'}{}^{r}C_{p}{}^{r}C_{r-s+q+q'-p+2}a^{r+q-p+1}b^{s-q+p-1}\nn
&&+\sum_{q=0}^{\lfloor s\rfloor}\sum_{q'=0}^{\lfloor s\rfloor-q}\,\sum_{p=0}^{r-s+q+q'}{}^{s-1}C_{q}{}^{s-1}C_{q'}{}^{r}C_{p}{}^{r}C_{r-s+q+q'-p}a^{r+q-p}b^{s-q+p}
\label{gamma21NS}
\eea
\bea
\g^{F(2)(2)}_{NS,rs}&=&-\sum_{q=0}^{\lfloor s\rfloor}\sum_{q'=0}^{\lfloor s\rfloor-q}\,\sum_{p=0}^{r+s-q-q'}{}^{s-1}C_q{}^{s-1}C_{q'}{}^{r}C_p{}^{r}C_{r+s-q-q'-p}a^{r+s-q-p}b^{q+p}\nn
&&\,\,+\sum_{q=0}^{\lfloor s\rfloor-2}\sum_{q'=0}^{\lfloor s\rfloor-q-2}\,\sum_{p=0}^{r+s-q-q'-2}{}^{s-1}C_q{}^{s-1}C_{q'}{}^{r}C_p{}^{r}C_{r+s-q-q'-p-2}a^{r+s-p-q-1}b^{q+p+1}.\nn
\label{gamma22NS}
\eea
Using the interchange symmetry $a \leftrightarrow b$, we find:
\bea
\g^{F(2)(2)}_{NS,rs}&=&\g^{F(1)(1)}_{NS,rs} \\
\g^{F(1)(2)}_{NS,rs}&=&\g^{F(2)(1)}_{NS,rs}.
\eea

Just as in the bosonic case, we shall now bring the $\g^{F(i)(j)}_{NS}$ coefficients into a form that involves only one summation over a pair of hypergeometric functions, binomial coefficients, and $a$ and $b$ terms raised to the appropriate powers. We present the computation for $i=j=1$ and simply give the results for the remaining cases. We start with:
\bea
\g^{F(1)(1)}_{NS,rs}&=&-\sum_{q=0}^{\lfloor s\rfloor}\sum_{q'=0}^{\lfloor s\rfloor-q}\,\sum_{p=0}^{r+s-q-q'}{}^{s-1}C_q{}^{s-1}C_{q'}{}^{r}C_p{}^{r}C_{r+s-q-q'-p}a^{q+p}b^{r+s-q-q'}\nn
&&+\sum_{q=0}^{\lfloor s\rfloor-2}\sum_{q'=0}^{\lfloor s\rfloor-q-2}\,\sum_{p=0}^{r+s-q-q'-2}{}^{s-1}C_q{}^{s-1}C_{q'}{}^{r}C_p{}^{r}C_{r+s-q-q'-p-2}a^{q+p+1}b^{r+s-p-q-1}.\nn
\eea
We now redefine our summation indices:
\bea
&&\text{Term 1: }\, j\equiv s-q-q'\implies q'=s-q-j\cr
&&\text{Term 2: }\, j\equiv s-q-q'-2\implies q'=s-q-j-2,
\label{redefintionsNS}
\eea
where in both cases $j$ is a positive half-integer. Since $q$ and $q'$ are both non-negative, we find:
\bea
&&\text{Term 1: }\,q'\geq 0~\implies~ q\leq s-j, \qquad\quad~~~\! q\geq 0~\implies~ j\leq s\nn
&&\text{Term 2: }\,q'\geq 0~\implies~ q\leq s-j-2, \qquad q\geq 0~\implies~ j\leq s-2.
\eea
This gives:
\bea
\g^{F(1)(1)}_{NS,rs}&=&-\sum_{j=1/2}^{s}b^{r+s}\left(\sum_{q=0}^{s-j}{}^{s-1}C_q{}^{s-1}C_{s-q-j}a^{q}b^{-q}\right)\left(\sum_{p=0}^{r+j}{}^{r}C_p{}^{r}C_{r+j-p}a^{p}b^{-p}\right) \cr
&&+\sum_{j=1/2}^{s-2}ab^{r+s-1}\left(\sum_{q=0}^{s-j-2}{}^{s-1}C_q{}^{s-1}C_{s-q-j-2}a^{q}b^{-q}\right)\left(\sum_{p=0}^{r+j}{}^{r}C_p{}^{r}C_{r+j-p}a^{p}b^{-p}\right).\nn
\eea
Evaluating the $q$ and $p$ sums in \textit{Mathematica}, we find:
\bea
\g^{F(1)(1)}_{NS,rs}&=&-\sum_{j=\h}^{s}{}^{s-1}C_{s-j}{}^{r}C_{r+j}b^{r+s} \nn
&&\quad{}\times{}_2F_1\left(1-s,j-s;j;\frac{a}{b}\right) {}_2F_1\left(-r,-j-r;1-j;\frac{a}{b}\right)\nn
&&{}+\sum_{j=\h}^{s-2}{}^{s-1/2}C_{s-j-2} {}^{r}C_{r+j}a b^{r+s-1}\nn
&&\quad{}\times {}_2F_1\left(1-s,j-s+2;j+2;\frac{a}{b}\right){}_2F_1\left(-r,-j-r;1-j;\frac{a}{b}\right).\nn
\eea

The other $\g^{F(i)(j)}_{NS,rs}$ coefficients are handled in the same manner. We present the final result along with the appropriate index redefinition that was made in each case.  For $\g^{F(1)(2)}_{NS,rs}$, we find:
\bea
\g^{F(1)(2)}_{NS,rs}&=&-\sum_{j=\h}^{s-2}{}^{s-1}C_{s-j-2}{}^{r}C_{r-j}a^{s-1}b^{r+1}\nn
&&{}\times_2F_1\left(-r,j-r;j;\frac{a}{b}\right)\,_2F_1\left(1-s,j-s+2;j+2;\frac{b}{a}\right)\nn
&&{}+\sum_{j=\h}^{s}{}^{s-1}C_{s-j}{}^{r}C_{r-j}a^{s}b^{r}\nn
&&{}\times_2F_1\left(-r,j-r;j+1;\frac{a}{b}\right)\,_2F_1\left(1-s,j-s;j;\frac{b}{a}\right),\nn
\eea
where in (\ref{gamma12NS}), we substituted:
\bea
&&\text{Term 1: }\,q'=s+j-q-2\nn
&&\text{Term 2: }\,q'=s+j-q.
\eea
For $\g^{F(2)(1)}_{NS,rs}$, we find:
\bea
\g^{F(2)(1)}_{NS,rs}&=&\sum_{j=\h}^{s}{}^{s-1}C_{s-j}{}^{r}C_{r-j}a^{r}b^{s}\nn
&&{}\times{}_2F_1\left(-r,j-r;j+1;\frac{b}{a}\right){}_2F_1\left(1-s,j-s;j;\frac{a}{b}\right)\nn
&&{}-\sum_{j=\h}^{s-2}{}^{s-1}C_{s-j-2}{}^{r}C_{r-j}a^{r+1}b^{s-1}\nn
&&{}\times {}_2F_1\left(-r-,j-r;j+1;\frac{b}{a}\right){}_2F_1\left(1-s,j-s+2;j+2;\frac{a}{b}\right) ,
{}   \nn\nn
\eea
where in (\ref{gamma21NS}), we substituted:
\bea
&&\text{Term 1: }\,q'=s-q-j\nn
&&\text{Term 2: }\,q'=s-q-j-2.
\eea
For $\g^{F(2)(2)}_{NS,rs}$, we find:
\bea
\g^{F(2)(2)}_{NS,rs}&=&\sum_{j=\h}^{s-2}{}^{s-1}C_{s-j-2}{}^{r}C_{r+j}a^{r+s-1}b\nn
&&{}\times _2F_1\left(-r,-j-r;1-j;\frac{b}{a}\right)\,_2F_1\left(1-s,j-s+2;j+2;\frac{b}{a}\right)\nn
&&{}-\sum_{j=\h}^{s}{}^{s-1}C_{s-j}{}^{r}C_{r+j}a^{r+s}\nn
&&{}\times  _2F_1\left(-r,-j-r;1-j;\frac{b}{a}\right)\,_2F_1\left(1-s,j-s;j;\frac{b}{a}\right),\nn
\eea
where in (\ref{gamma22NS}), we substituted:
\bea
&&\text{Term 1: }\,q'=s+j-q-2\cr
&&\text{Term 2: }\,q'=s+j-q.
\eea

\section{Computation of $\g^{F(i)(j)}_{R+-}$}\label{Ramond}
Here we perform the computation of the $\g^{F(i)(j)}_{R+-,mn}$ coefficients, where the $R+-$ indicates that we are capping the squeezed state with the positive Ramond vacuum on copy $1$ and the negative Ramond vacuum on copy $2$. In order to obtain an empty t-plane from this capping state we must apply one additional spectral flow beyond those performed in the $NS$ case. We then expand the resulting modes in terms of modes natural to the $t$ plane in order to compute $\g^{F(i)(j)}_{R+-,mn}$.

\subsection{Modes in t-plane from additional spectral flow by $\a=-1$}
Here we obtain the $t$-plane modes after one additional spectral of $\a=-1$ around the point $t=0$. We remind the reader that for the $NS$ sector computation, we performed four spectral flows:
\bea
&&\a=1\text{ around }t=-b\nn
&&\a=1\text{ around }t=-a\nn
&&\a=-1\text{ around }t=-\sqrt{ab}\nn
&&\a=-1\text{ around }t=\sqrt{ab}.
\eea
The resulting modes were given by
\bea
d'^{(1)f,+A}_{r}&=&\frac{1}{2\pi i}\int_{t=\infty}\psi^{(1)f,+A}(t)t^{-r-1/2}(t^{2}-ab)(t+a)^{r-1}(t+b)^{r-1}\diff t\nn
d'^{(1)f,-A}_{r}&=&\frac{1}{2\pi i}\int_{t=\infty}\psi^{(1)f,-A}(t)t^{-r-1/2}(t+a)^{r}(t+b)^{r}\diff t\nn
d'^{(2)f,+A}_{r}&=&-\frac{1}{2\pi i}\int_{t=0}\psi^{(2)f,+A}(t)t^{-r-1/2}(t^{2}-ab)(t+a)^{r-1}(t+b)^{r-1}\diff t\nn
d'^{(2)f,-A}_{r}&=&-\frac{1}{2\pi i}\int_{t=0}\psi^{(2)f,-A}(t)t^{-r-1/2}(t+a)^{r}(t+b)^{r}\diff t.
\eea
We now apply one additional spectral flow of $\a=-1$ around the point, $t=0$. We also note that the Ramond sector has integer-indexed fermions, so we will switch from the half-integer mode index $r$ to the integer index $n$.

Under this last spectral flow, the fermion field changes as follows:
\bea
\psi^{(i)f,\pm A}(t)\to t^{\pm 1/2}\psi^{(i)f,\pm}(t).
\eea
The modes then become
\bea
d'^{(1)f,+A}_{n}&\to&\hat{d}^{(1)f,+A}_{n}~=~\frac{1}{2\pi i}\int_{t=\infty}\psi^{(1)f,+A}(t)t^{-n}(t^{2}-ab)(t+a)^{n-1}(t+b)^{n-1}\diff t\nn
d'^{(1)f,-A}_{n}&\to&\hat{d}^{(1)f,-A}_{n}~=~\frac{1}{2\pi i}\int_{t=\infty}\psi^{(1)f,-A}(t)t^{-n-1}(t+a)^{n}(t+b)^{n}\diff t\nn
d'^{(2)f,+A}_{n}&\to&\hat{d}^{(2)f,+A}_{n}~=~-\frac{1}{2\pi i}\int_{t=0}\psi^{(2)f,+A}(t)t^{-n}(t^{2}-ab)(t+a)^{n-1}(t+b)^{n-1}\diff t\nn
d'^{(2)f,-A}_{n}&\to&\hat{d}^{(2)f,-A}_{n}~=~-\frac{1}{2\pi i}\int_{t=0}\psi^{(2)f,-A}(t)t^{-n-1}(t+a)^{n}(t+b)^{n}\diff t.\nn
\label{Rsector_tmodes}
\eea 

We now expand the modes (\ref{Rsector_tmodes}) in terms of modes natural to the $t$ plane.  Since the additional spectral flow produced a factor of $t^{\pm\h}$, the binomials we wish to expand are identical to the ones found in the NS case.  We thus present only the results: 
\bea
\hat{d}_{n}^{(1)f,+A}&=&\sum_{j,j'\geq 0} {}^{n-1}C_j{}^{n-1}C_{j'} a^j b^{j'} \tilde{d}^{+A}_{n-j-j'+1/2}\nn
&&{}-\sum_{j,j'\geq 0} {}^{n-1}C_j{}^{n-1}C_{j'} a^{j+1} b^{j'+1}\tilde{d}_{n-j-j'-3/2}^{+A}\nn
\hat{d}_{n}^{(1)f,-A}&=&\sum_{j,j'\geq 0} {}^{n}C_j{}^{n}C_{j'} a^{j} b^{j'}\tilde{d}^{-A}_{n-j-j'-1/2}\nn
\hat{d}_{n}^{(2)f,+A}&=&-\sum_{j,j'\geq 0}{}^{n-1}C_j{}^{n-1}C_{j'} a^{n-j-1} b^{n-j'-1} \tilde{d}^{+A}_{j+j'-n+5/2}\nn
&&{}+\sum_{j,j'\geq 0} {}^{n-1}C_j{}^{n-1}C_{j'} a^{n-j} b^{n-j'}\tilde{d}^{+A}_{j+j'-n+1/2}\nn
\hat{d}_{n}^{(2)f,-A}&=&-\sum_{j,j'\geq 0} {}^{n}C_j{}^{n}C_{j'} a^{n-j} b^{n-j'}\tilde{d}^{-A}_{j+j'-n-1/2}.\nn
\label{tplanemodesNS}
\eea

\subsection{Computing $\g^{F(i)(j)}_{R+-,mn}$}
Here we show the computation of the $\g^{F(i)(j)}_{R+-}$ coefficients.
We shall explicitly show the computation for the case $i=j=1$. The other cases are computed in the same way and we shall simply state their results. Using (\ref{FermionRCoverRatio1}) and (\ref{tplanemodesNS}), we find:
\bea
\g^{F(1)(1)}_{R+-,mn}&=&\sum_{j,j',k,k'\geq 0}{}^{n-1}C_k{}^{n-1}C_{k'}{}^{m}C_j{}^{m}C_{j'}a^{k+j} b^{k'+j'}~\frac{\nstbra\tilde{d}^{++}_{n-k-k'+1/2}\tilde{d}^{--}_{m-j-j'-1/2}\nstket}{\nstbra\nstclose_{t}}\nn
&&{}-\sum_{j,j',k,k'\geq 0}{}^{n-1}C_k{}^{n-1}C_{k'}  {}^{m}C_j{}^{m}C_{j'}a^{k+j+1} b^{k'+j'+1}\nn
&&\quad{}\times\frac{{}_{t}\langle 0_{NS}|\tilde{d}_{n-k-k'-3/2}^{++}\tilde{d}^{--}_{m-j-j'-1/2}|0_{NS}\rangle_{t}}{\nstbra\nstclose_{t}}.\nn
\label{gamma11R+-}
\eea
Looking at the first term, it is clear that the summand vanishes except when:
\bea
m+n-j-j'-k-k'=0~&\implies&~ j'=m+n-k-k'-j\nn
n-k-k'+\h > 0 ~&\implies&~ n-k \geq k'.
\eea
Furthermore,
\bea
k' \geq 0 ~&\implies&~ n \geq k\nn
j' \geq 0 ~&\implies&~ m+n-k-k' \geq j.
\eea
We can thus re-write (\ref{gamma11R+-}) as:
\bea
\g^{F(1)(1)}_{R+-,mn}&=&\sum_{k=0}^{n}\sum_{k'=0}^{n-k}\,\sum_{j=0}^{m+n-k-k'}{}^{n-1}C_k{}^{n-1}C_{k'}{}^{m}C_j{}^{m}C_{m+n-k-k'-j}a^{k+j}b^{m+n-j-k}\nn
&&{}-\sum_{j,j',k,k'\geq 0}{}^{n-1}C_k{}^{n-1}C_{k'}  {}^{m}C_j{}^{m}C_{j'}a^{k+j+1}b^{k'+j'+1}\nn
&&\quad{}\times\frac{{}_{t}\langle 0_{NS}|\tilde{d}_{n-k-k'-3/2}^{++}\tilde{d}^{--}_{m-j-j'-1/2}|0_{NS}\rangle_{t}}{\nstbra\nstclose}.\nn
\label{gamma11R+-_2}
\eea
Applying the same line of reasoning for the second term, we see:
\bea
m+n-j-j'-k-k'-2=0 ~&\implies&~ j'=m+n-k-k'-j-2 \nn
n-k-k'-{3\over 2} > 0 ~&\implies&~ n +2-k \geq k' \nn
j'\geq 0 ~&\implies& ~ m+n-k-k'-2\geq j.\nn
\eea
So (\ref{gamma11R+-_2}) becomes
\bea
\g^{F(1)(1)}_{R+-,mn} &=&-\sum_{k=0}^{n}\sum_{k'=0}^{n-k}\,\sum_{j=0}^{m+n-k-k'}{}^{n-1}C_k{}^{n-1}C_{k'} {}^{m}C_j{}^{m}C_{m+n-k-k'-j}a^{j+k} b^{m+n-j-k}\nn
&&{}+\sum_{k=0}^{n-2}\sum_{k'=0}^{n-k-2}\,\sum_{j=0}^{m+n-k-k'-2}{}^{n-1}C_k{}^{n-1}C_{k'}{}^{m}C_j{}^{m}C_{m+n-k-k'-j-2}a^{j+k+1}b^{m+n-j-k-1}.\nn\label{gamma11R+-_f}
\eea

Following these same steps for the other cases, we obtain:
\bea
\g^{F(1)(2)}_{R+-,mn}&=&-\sum_{k=0}^{n-3}\sum_{k'= 0}^{n-k-3}\,\sum_{j= 0}^{m-n+k+k'+2}{}^{n-1}C_{k}{}^{n-1}C_{k'}{}^{m}C_{j}{}^{m}C_{m-n+k+k'-j+2}a^{n+j-k-1}b^{m-j+k+1}\nn
&&{}+\sum_{k= 0}^{n-1}\sum_{k'= 0}^{n-k-1}\,\sum_{j=0}^{m-n+k+k'}{}^{n-1}C_{k}{}{}^{n-1}C_{k'}{}^{m}C_{j}{}^{m}C_{m-n+k+k'-j}a^{n+j-k}b^{m-j+k}\nn\label{gamma12R+-}
\eea
\bea
\gamma_{R+-,mn}^{F(2)(1)}&=&\sum_{k=0}^{n}\sum_{k'=0}^{n-k}\sum_{j=0}^{m-n+k+k'}{}^{n-1}C_{k}{}^{n-1}C_{k'}{}^{m}C_{j}{}^{m}C_{m-n+k+k'-j}a^{m-j+k}b^{n+j-k}\nn
&&{}-\sum_{k=0}^{n-2}\sum_{k'=0}^{n-k-2}\,\sum_{j=0}^{m-n+k+k'+2}{}^{n-1}C_{k}{}^{n-1}C_{k'}{}^{m}C_{j}{}^{m}C_{m-n+k+k'-j+2}a^{m-j+k+1}b^{n+j-k-1}\nn\label{gamma21R+-}
\eea
\bea
\g^{F(2)(2)}_{R+-,mn}&=&\sum_{k=0}^{n-3}\sum_{k'=0}^{n-k-3}\,\sum_{j=0}^{m+n-k-k'-2}{}^{n-1}C_{k}{}^{n-1}C_{k'}{}^{m}C_{j}{}^{m}C_{m+n-k-k'-j-2}a^{m+n-j-k-1}b^{j+k+1}\nn
&&{}-\sum_{k=0}^{n-1}\sum_{k'=0}^{n-k-1}\,\sum_{j=0}^{m+n-k-k'}{}^{n-1}C_{k}{}^{n-1}C_{k'}{}^{m}C_{j}{}^{m}C_{m+n-k-k'-j}a^{m+n-j-k}b^{j+k}.\nn\label{gamma22R+-}
\eea
The physics is symmetric under Copy 1 $\leftrightarrow$ Copy 2.  This symmetry manifests in the following relations:
\bea
\g^{F(1)(1)}_{R+-;m,0}&=&-\g^{(2)(1)}_{R+-;m,0},\quad m>1\nn
\g^{F(1)(1)}_{R+-,mn}&=&\g^{F(2)(2)}_{R+-,mn},\quad m,n > 1\nn
\g^{F(1)(2)}_{R+-,mn}&=&\g^{F(2)(1)}_{R+-,mn},\quad m,n > 1 \nn
\g^{(2)(1)}_{R+-;0,n} &=& \d_{n,0} \nn
\g^{(2)(2)}_{R+-;0,n} &=& 0. \label{symmetryrequirements}
\eea
The manner in which these relations enforce the copy symmetry is examined in detail in Appendix \ref{CopySymmetry}. The fact that these relations are satisfied will become evident when we re-express the coefficients in terms of hypergeometric functions. As usual we present the computation for $i=j=1$ and simply give the results for the remaining cases.

The symmetry requirements of (\ref{symmetryrequirements}) suggests that $\g^{F(1)(1)}_{R+-,mn}$ may behave differently for $n=0$. Indeed, it is convenient to treat the $n=0$ case separately.  Setting $n=0$ in (\ref{gamma11R+-_f}), we find:
\bea
\g^{F(1)(1)}_{R+-;m,0} &=& -\sum_{j=0}^m a^j b^{m-j} \,{}^m C_j {}^m C_{m-j} ~=~ -b^m {}_2 F_1 \left ( -m,-m,1,{a\over b}\right).
\eea
When $n>0$, we begin:
\bea
&&\left[\g^{F(1)(1)}_{R+-,mn}\right]_{n > 0}\nn
&&\quad = -\sum_{k=0}^{n}\sum_{k'=0}^{n-k}\,\sum_{j=0}^{m+n-k-k'}{}^{n-1}C_k{}^{n-1}C_{k'} {}^{m}C_j{}^{m}C_{m+n-k-k'-j}a^{j+k}b^{m+n-j-k}\nn
&&\quad \quad +\sum_{k=0}^{n-2}\sum_{k'=0}^{n-k-2}\,\sum_{j=0}^{m+n-k-k'-2}{}^{n-1}C_k{}^{n-1}C_{k'}{}^{m}C_j{}^{m}C_{m+n-k-k'-j-2}a^{j+k+1} b^{m+n-j-k-1}.\nn
\eea
We now make the following index redefinitions:
\bea
&&\text{Term 1: }\,p\equiv n-k-k'+1/2\implies k'=n-k-p+1/2\nn
&&\text{Term 2: }\,p\equiv n-k-k'-3/2\implies k'=n-k-p-3/2,
\label{redefintionsR+-}
\eea
where in both cases $p$ is a positive half-integer.  We then find:
\bea
&&\text{Term 1}:\quad k'\geq 0\implies k\leq n-p+1/2~~~~\text{and}~~~~k,k'\geq 0\implies p\leq n+1/2\nn
&&\text{Term 2}:\quad k'\geq 0\implies k\leq n-p-3/2~~~~\text{and}~~~~k,k'\geq 0\implies p\leq n-3/2.\nn
\eea
This gives:
\bea
\left[\g^{F(1)(1)}_{R+-,mn}\right]_{n > 0} &=&-\sum_{p=1/2}^{n+1/2}b^{m+n}\left(\sum_{k=0}^{n-p+1/2}{}^{n-1}C_k{}^{n-1}C_{n-k-p+1/2}a^{k}b^{-k}\right)\nn
&&\qquad {}\times \left(\sum_{j=0}^{m+p-1/2}{}^{m}C_j{}^{m}C_{m+p-j-1/2}a^{j}b^{-j}\right)\nn
&&{}+\sum_{p=1/2}^{n-3/2}ab^{m+n-1}\left(\sum_{k=0}^{n-p-3/2}{}^{n-1}C_k{}^{n-1}C_{n-k-p-3/2}a^{k}b^{-k}\right)\nn
&&\qquad {}\times \left(\sum_{j=0}^{m+p-1/2}{}^{m}C_j{}^{m}C_{m+p-j-1/2}a^{j} b^{-j}\right).\nn\label{hypergeometricintermediate}
\eea
Careful examination of (\ref{hypergeometricintermediate}) reveals that the $p=\h$ terms either independantly vanish or precisely cancel for all values of $n$.  We thus shift the lower bound of the sums to $p = \tfrac{3}{2}$.  Performing the $k$ and $j$ sums in $\textit{Mathematica}$, we are left with:
\bea
\left[\g^{F(1)(1)}_{R+-,mn}\right]_{n>0}&=& -\sum_{p = {3\over 2}}^{n+\h}{}^{n-1}C_{n-p+\h}{}^m C_{p-\h}a^{p-\h}b^{m+n-p+\h}\nn
&&\quad{}\times {}_2 F_1 \left ( 1-n, p-n-\h; p-\h;{a\over b}\right ) {}_2 F_1 \left ( p-m-\h,-m;p+\h;{a\over b}\right ) \nn
&&{}+\sum_{p = {3\over 2}}^{n-{3\over2}}{}^{n-1}C_{n-p-{3\over2}}{}^m C_{p-\h}a^{p+\h}b^{m+n-p-\h}\nn
&&\quad{}\times {}_2 F_1 \left( 1-n, p-n+{3\over2}; p+{3\over2};{a\over b}\right) {}_2 F_1 \left( p-m-\h,-m;p+\h;{a\over b}\right).\nn
\eea

The other cases are performed similarly, and we present only the results along with the index redefinitions used.
\bea
\g^{F(1)(2)}_{R+-,mn}&=&-\sum_{p=\h}^{\min\left(m-\h,n-{5\over2}\right)}{}^{n-1}C_{n-p-{5\over2}}{}^{m}C_{m-p-\h}a^{n-1} b^{m+1}\nn
&&{}_2F_1\left(p-m+\h,-m;p+{3\over2};\frac{a}{b}\right) {}_2F_1\left(1-n,p-n+{5\over2};p+{5\over2};\frac{b}{a}\right)\nn
&&{}+\sum_{p=\h}^{\min\left(m-\h,n-\h\right)}{}^{n-1}C_{n-p-\h}{}^{m}C_{m-p-\h}a^{n}b^{m}\nn
&&{}\times {}_2F_1\left(p-m+\h,-m;p+{3\over2};\frac{a}{b}\right) \,\, \,\,_2F_1\left(1-n,p-n+\h;p+\h;\frac{b}{a}\right),\nn
\label{gammaF12}
\eea
where in (\ref{gamma12R+-}), we substituted:
\bea
&&\text{Term 1: }\,k'\equiv n+p-k-{5\over2} \nn
&&\text{Term 2: }\,k'\equiv n+p-k-\h.
\eea
\bea
\left[\g^{F(2)(1)}_{R+-,mn}\right]_{n=0}&=& \sum_{j=0}^m{}^m C_j{}^m C_{m-j}a^{m-j}b^j~=~a^m{}_2F_1\left(-m,-m,1,{b\over a}\right)\nn
\left[\g^{F(2)(1)}_{R+-,mn}\right]_{n>0}&=& \sum_{p={1\over 2}}^{\min\left(m-\h,n-\h\right)}{}^{n-1}C_{n-p-\h}{}^m C_{m-p-\h}a^m b^n\nn
&&{}\times {}_2 F_1 \left(1-n,p-n+\h;p+\h;{a\over b}\right){}_2 F_1 \left(p-m+\h,-m;p+\frac{3}{2};{b\over a}\right) \nn
&&{}-\sum_{p={1\over2}}^{\min\left(m-\h,n-{5\over2}\right)}{}^{n-1}C_{n-p-{5\over2}}{}^m C_{m-p-\h}a^{m+1}b^{n-1}\nn
&&{}\times {}_2 F_1 \left(1-n,p-n+{5\over2};p+{5\over2};{a\over b}\right){}_2 F_1 \left(p-m+\h,-m;p+{3\over2};{b\over a}\right),\nn
\label{gammaF21}
\eea
where in (\ref{gamma21R+-}), we substituted:
\bea
&&\text{Term 1: }\,k'\equiv n-p-k-\h\nn
&&\text{Term 2: }\,k'\equiv n-p-k-{5\over 2}.
\eea
\bea
\g^{F(2)(2)}_{R+-,mn} &=& \sum_{p={3\over2}}^{n+\h}{}^{n-1}C_{p-{3\over 2}}{}^mC_{m-p+\h}a^{m+n-p+\h}b^{p-\h}\nn
&&{}\times{}_2F_1\left(1-n,p-n-\h;p-\h;{b\over a}\right)\,{}_2F_1\left(p-m-\h,-m;p+\h;{b\over a}\right)\nn
&&{}- \sum_{p={3\over 2}}^{n-\frac{3}{2}}{}^{n-1}C_{p+\h}{}^mC_{m-p+\h}a^{m+n-p-\h}b^{p+\h}\nn
&&{}\times{}_2F_1\left(1-n,p-n+{3\over2};p+{3\over2};{b\over a}\right) {}_2F_1\left(p-m-\h,-m;p+\h;{b\over a}\right),\nn
\eea
where in (\ref{gamma22R+-}), we substituted:
\bea
&&\text{Term 1: }\,k'\equiv n+p-k-5/2\nn
&&\text{Term 2: }\,k'\equiv n+p-k-1/2.
\eea
Using the $a\leftrightarrow b$ symmetry, it is clear that the relations (\ref{symmetryrequirements}) are satisfied.  We also show in Appendix \ref{GammaRelations} the additional relationship:
\bea
\left[\g^{F(1)(1)}_{R+-,mn}\right]_{n>0} &=& -\left[\g^{F(2)(1)}_{R+-,mn}\right]_{m,n>0}.
\eea
We thus have only a single linearly independent fermion coefficient for the non-zero modes (along with coefficients for zero modes).

\section{NS vs R sectors in the final state}\label{NS-R Sectors}

We have started with two singly wound component strings, both in the R sector. This sector was chosen because the physical D1D5 black hole has R periodicity for its fermions. We applied two twists to this initial state, and then tried to determine the final state by taking an inner product with all possible states.

Here we found the following. We can get a nonzero inner product by taking a final state which has two singly wound strings, each in the R sector. But we  also find a nonzero inner product if we take a final state which has two singly wound strings, each in the NS sector. Does this mean that the initial R sector state can evolve to a state in the NS sector?

Such an evolution would be surprising, since it is known that all states of the D1D5 system can be accounted for by the different choices of R sector states. So why are we finding a nonzero inner product with NS sector states? To understand this, we consider an analogy: the example of the Ising model.

 \begin{figure}[htbp]
\begin{center}
\includegraphics[scale=.52]{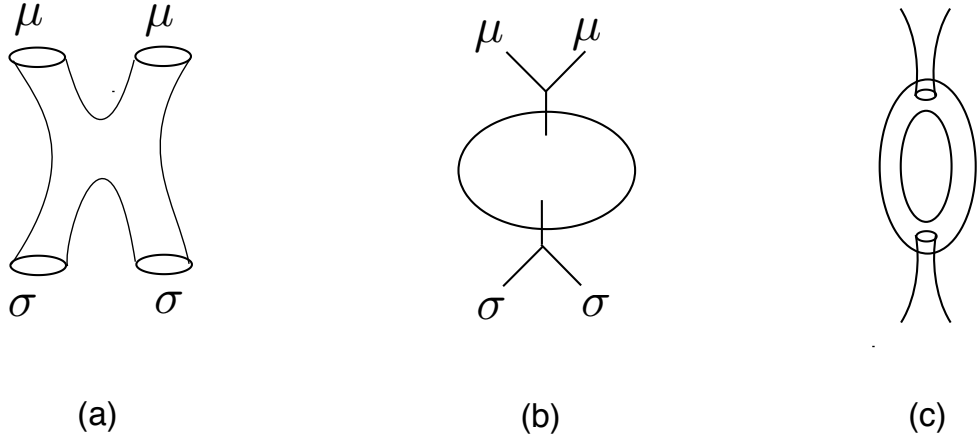}
\caption{{(a) An amplitude in the Ising model where we have put $\sigma$ states on two legs and $\mu$ states on the other two. (b) The same amplitude drawn differently, using conformal invariance. (c) A 1-loop amplitude; only a single set of complete fields should run in the loop.}}
\label{amplitude}
\end{center}
\end{figure}

The Ising model has 3 primary operators: $I, \sigma, \epsilon$. However it has a dual representation as well, where the operator order parameter $\sigma$ is replaced by the disorder parameter $\mu$. 

Let us use $\sigma$ as our basic variable; not the dual variable $\mu$. Now consider the CFT  amplitude shown in fig.\ref{amplitude}(a). The initial state has two copies of the CFT on a circle, each created by the application of a $\sigma$ operator to the identity. The final state again has two copies of the CFT on a circle, but each state is generated by the application of $\mu$ to the identity state. Is this transition amplitude nonzero?

We can redraw this amplitude in the form shown in fig.\ref{amplitude}(b). Now it looks like the amplitude should be nonzero" the two $\sigma$ operators fuse to the identity, and the two $\mu$ operators do as well. The insertion of two identity operators into the sphere gives a nonzero amplitude. 

So should we conclude that an initial state made of two $\sigma$ operators can evolve to a state containing two $\mu$ operators? If this were true, then we would have to allow $\mu$ operators to run in loops like the one shown in fig.\ref{amplitude}(c), along with the $\sigma$ operator that can run in the loop. 

But at this stage we note that this looks like double counting: a complete set of operators is given by $I, \sigma, \epsilon$, without the inclusion of $\mu$. The operator $\mu$ does not need to be included in loops. This shows up in the fact that $\mu$ is not local with respect to $\sigma$. 

In a similar way we note that the NS and R sectors are different. The R sector state is created by applying a spin field to the NS sector state, and this spin field is not local with respect to the fermions in the theory. We can start with the NS sector, and stay in this sector all along, or we can spectral flow to the R sector, and stay in the R sector. But we should not include both sectors. The formalism we have used started with the states in the R sector, thus we should keep only the inner products we computed for transitions to the R sector, and not the ones for the NS sector.

\section{Analysis}\label{Analysis}

In the single twist case, we were able to write down explicit closed form expressions for $\gamma^B_{mn}, \gamma^F_{mn}$. In the 2-twist case, 
we do not have such closed form expressions.  In the single twist case the dependence on the position of the twist was trivial; it was just an overall phase. In the 2-twist case we expect a nontrivial dependence on the  separation between the two twists $\Delta w$. We would also like to see how the $\gamma^B_{mn}, \gamma^F_{mn}$ fall off for large values of $m,n$. While it is not possible to see these features from the form of these functions as finite sums, numerical analysis turns out to reveal the information we seek. In this section we will explore the functions $\gamma^B_{mn}, \gamma^F_{mn}$ by computing them using Mathematica.

\subsection{Wick Rotation and Origin Choice}

Note that the physical problem of the black hole needs us to do all computations for Minkowski signature on the cylinder. Thus we first rotate Euclidean time to Minkowski time. 
\bea
\t \to it ~\implies~ w = it + i\s.
\eea
Here $w$ is purely imaginary.

Now that we have returned to Minkowski time, we make the convenient choice of setting our origin midway between $w_1$ and $w_2$.  We thus have:
\bea
t_2 = -t_1 = {\D t \over 2}, \quad \s_2 = -\s_1 = {\D\s\over 2}.
\eea
With the choice of coordinates, we make several  observations that will simplify our algebra:
\bea
z_2 &=& e^{w_1} ~=~ e^{i{\D t + \D s \over 2}} ~=~ a+b + 2\sqrt{ab}\nn
z_1 &=& e^{w_1} ~=~ e^{-i{\D t + \D s \over 2}} ~=~ a+b - 2\sqrt{ab},
\eea
from which we find:
\bea
a+b &=& {z_2 + z_1 \over 2}\nn
\sqrt{ab} &=& {z_2 - z_1 \over 4},
\eea
and thus\footnote{We choose to have $b\to0$ as $\D w\to0$ so that the initial and final Copy 2 strands have the same image in this limit.}:
\bea
\sqrt{a} &=& \h\left(\sqrt{z_2} + \sqrt{z_1}\right) ~=~ \cos\left( {\D t + \D \s \over 4}\right)\nn
\sqrt{b} &=& \h\left(\sqrt{z_2} - \sqrt{z_1}\right) ~=~ i\sin\left( {\D t + \D \s \over 4}\right).\nn
\eea
This gives
\bea
a &=& \cos^2\left ({\D w \over 4i}\right)\nn
b &=& -\sin^2\left ({\D w \over 4i}\right) ~=~ 1-a,
\eea
eliminating the parameter $b$.  This also makes it clear that our $\g$ parameters will be periodic in $\D w$, with a period of $4\pi i$.

\subsection{Numerical Analysis}
Now that we have reduced our $\g$ coefficients to a single spacetime parameter, we can numerically calculate their values for various choices of $m$, $n$, and $\D w$.  We present here several figures plotting the behavior of the bosonic and fermionic $\g$ coefficients as we vary these parameters.

\begin{figure}
\includegraphics[width=0.5\columnwidth]{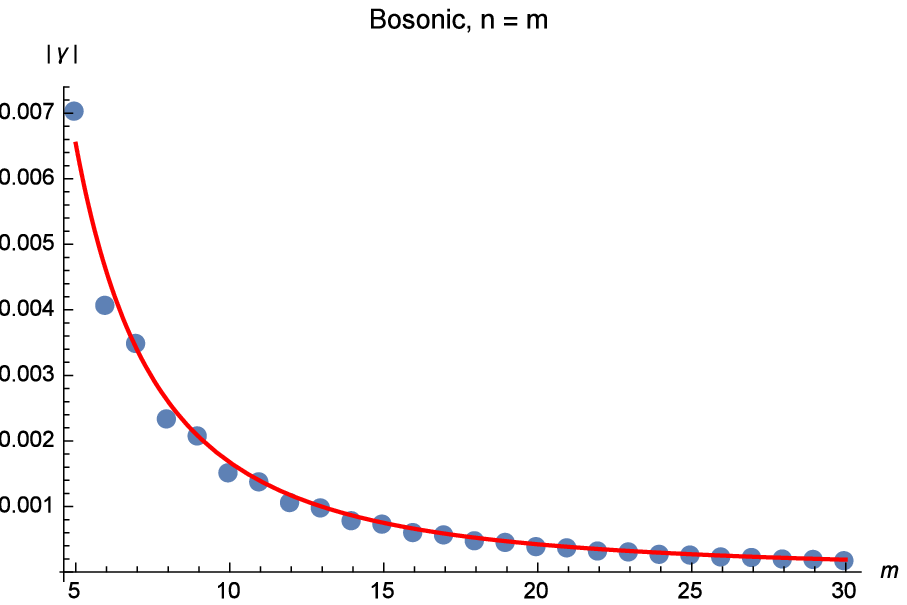}
\includegraphics[width=0.5\columnwidth]{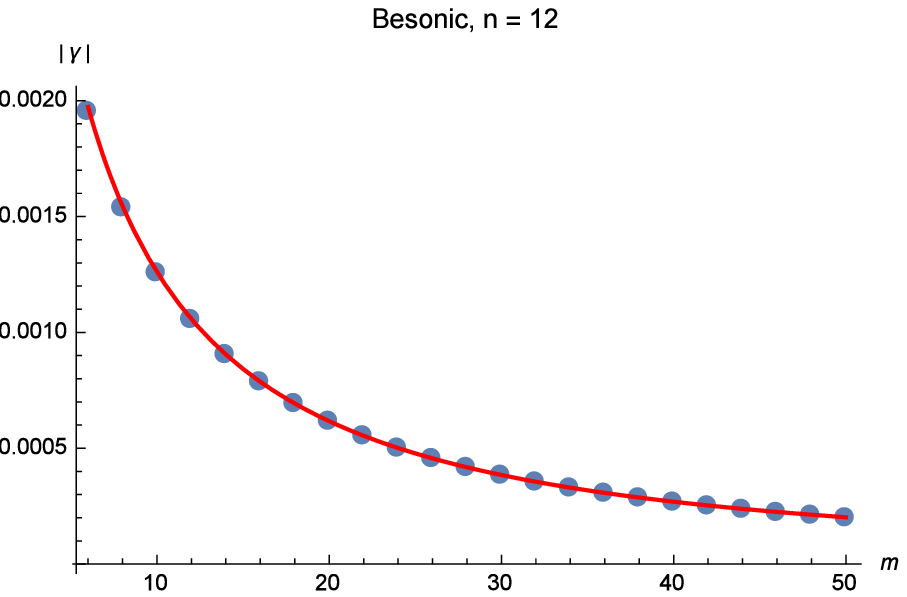}
\caption{Numerical values of $\left | \g^{B,(1)(1)}_{mn}\right | $ for $\D w = i\pi$.  For $n=m$ (left), a fit of the form $\G^2(m+\tfrac{1}{2})/[m\G^2(m+1)]$ is shown.  For $n=12$ (right), a fit of the form $\G(m+\tfrac{1}{2})/[\G(m+1)(m+12)]$ is shown.  The coefficient clearly behaves as $\G(m+\tfrac{1}{2})\G(n+\tfrac{1}{2})/[\G(m+1)\G(n+1)(m+n)]$.  Both plots show only $m>4$, as these simple fits do not work well for very low values of $m$ and $n$.  The plot on the right shows only even $m$, as the coefficient vanishes at $\D w = i\pi$ when $m+n$ is odd.}
\label{Bosonic_Fixed_w}
\end{figure}

\begin{figure}
\includegraphics[width=0.5\columnwidth]{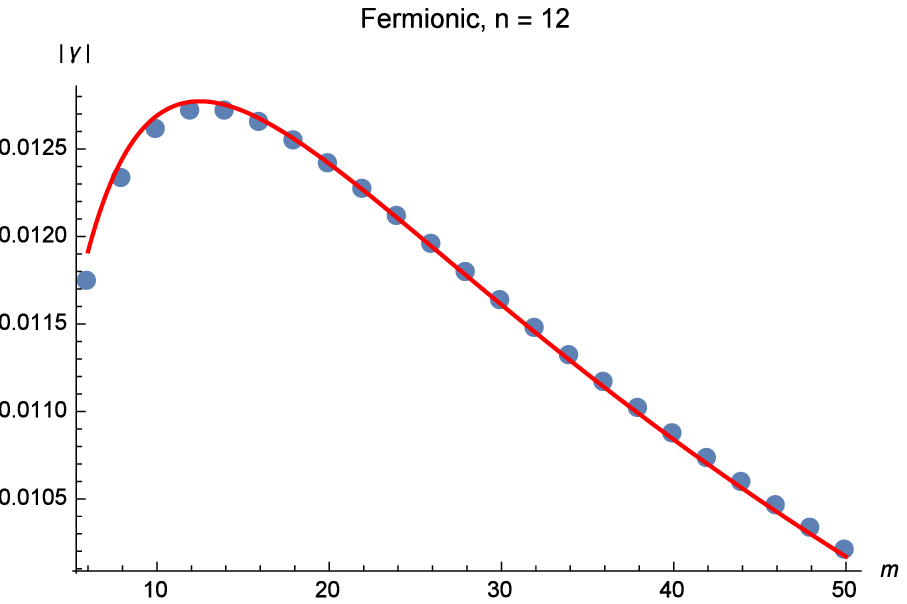}
\includegraphics[width=0.5\columnwidth]{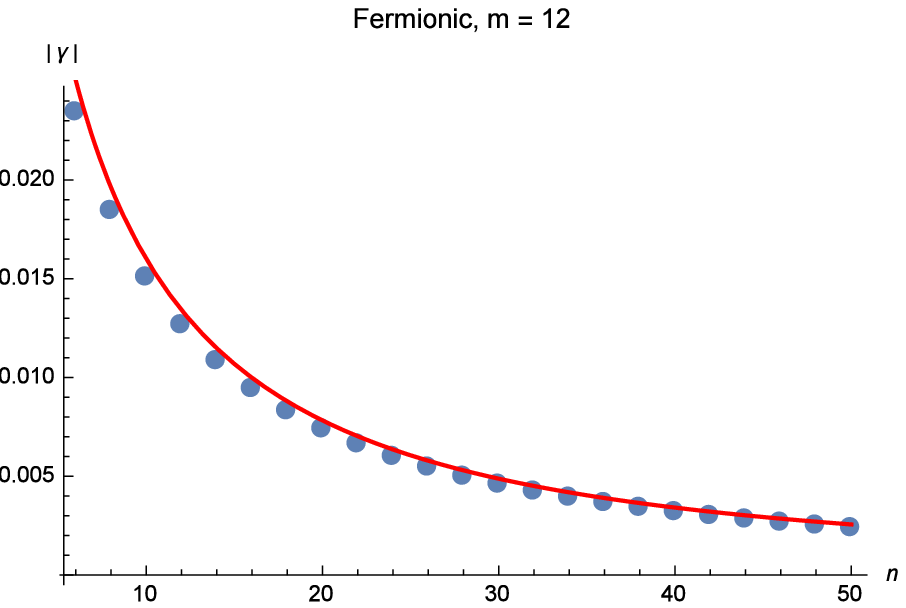}
\caption{Numerical values of $\left | \g^{F,(1)(1)}_{mn}\right | $ for $\D w = i\pi$.  For $n=12$ (left), a fit of the form $\G(m+\tfrac{1}{2})/[\G(m)(m+12)$ is shown.  For $m=12$ (right), a fit of the form $\G(n+\tfrac{1}{2})/[\G(n+1)(n+12)]$ is shown.  The coefficient clearly behaves as $\G(m+\tfrac{1}{2})\G(n+\tfrac{1}{2})/[\G(m)\G(n+1)(m+n)]$.  Both plots show only even $m,n>4$, as these simple fits do not work well for very low values of $m$ and $n$ and this coefficient also vanishes at $\D w = i\pi$ when $m+n$ is odd.}
\label{Fermionic_Fixed_w}
\end{figure}

Figures \ref{Bosonic_Fixed_w} and \ref{Fermionic_Fixed_w} show the behavior of the bosonic and fermionic $\g^{(1)(1)}_{mn}$ as $m$ or $n$ are varied.  The value of $\Delta w$ is held fixed. By considering a set of such graphs, ranging over different values of $\Delta w$, we can conjecture a simple approximate dependence of the $\gamma$ on $m,n$:
\bea
\g^{B,(1)(1)}_{mn} &\sim& {1\over (m+n)}{\G(m+\h)\G(n+\h)\over \G(m+1)\G(n+1)}\nn
\g^{F,(1)(1)}_{mn} &\sim& {1\over (m+n)}{\G(m+\h)\G(n+\h)\over \G(m)\G(n+1)}.
\eea
This leads us to guess a simple relationship for the magnitudes of the bosonic and fermionic $\g$ coefficients.  Fixing the sign, we have:
\bea
\g^{F,(1)(1)}_{mn} &=& -m\g^{B,(1)(1)}_{mn}.\label{GBFRelation}
\eea
We prove this relationship analytically in Appendix \ref{boson fermion relation}.  The relationship also exists for the first-order $\g$ coefficients, though this was not noted in \cite{acm1}.

For large $m$ and $n$, (\ref{GBFRelation}) simplifies to:
\bea
\g^{B,(1)(1)}_{mn} &\sim& {1\over \sqrt{mn}(m+n)}\nn
\g^{F,(1)(1)}_{mn} &\sim& \sqrt{m\over n}{1\over (m+n)},
\eea
which is exactly the same behavior as the single-twist $\g$.  Thus for any specific separation, each two-twist $\g$ looks just like a re-scaled single-twist $\g$.

\begin{figure}[h]
\includegraphics[width=0.5\columnwidth]{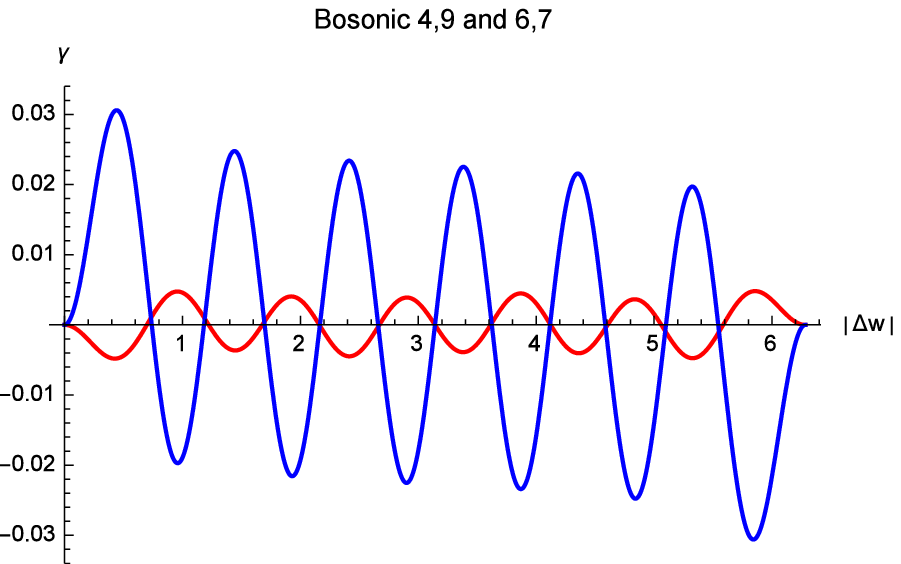}
\includegraphics[width=0.5\columnwidth]{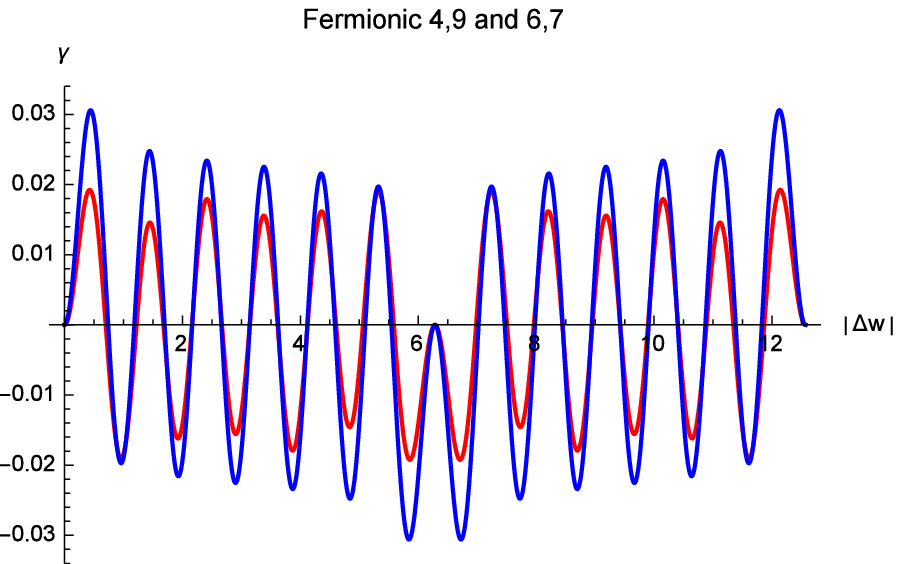}
\caption{Numerical values of $\left | \g^{F,(1)(1)}_{mn}\right | $ for $m=4,n=9$ (red) and $m=6,n=7$ (blue).  The $\D w$ dependence is clearly an oscillatory function.  The number of peaks is determined by $m+n$.}
\label{wPlot}
\end{figure}

Next we fix values for $m$ and $n$ while varying $\D w$.  As seen in Figure \ref{wPlot}, this analysis reveals that the $\D w$ dependance is an oscillation with a  frequency set by $m+n$. The amplitude of the oscillations, however does not remain constant over the full range of $\Delta w$. 

Combined with the previous result we reach a remarkable conclusion: The two-twist $\g$ coefficients have the exact same large $m,n$ behavior as their one-twist counterparts, except for the addition of an oscillation in $\D w$.

\section{Discussion}

Our goal is to see thermalization in the D1D5 CFT. This thermalization is expected to be dual to the process of black hole formation in the gravity theory. Thus if can understand thermalization in the CFT, we get a window on the some of the deepest questions associated to black holes. 

The free D1D5 CFT itself has been very useful; it gives correctly the extremal and near-extremal entropies of the black hole, as well as the greybody factors of radiation from the near-extremal hole. But the free theory cannot show thermalization, since the excitations on the D1D5 brane system are free fields at the orbifold point. To reach the supergravity point, we have to consider the deformation operator $O$ that takes us away from the orbifold point.

In our earlier studies of this problem, we had worked out the effect of one insertion of the deformation operator. Such an insertion takes two singly wound component strings, for example, to one doubly wound component string. Though this is a nontrivial interaction, creating excitations on the doubly wound string, it is hard to see evidence of thermalization at this level. One reason is that the vacuum state of the doubly wound string is different from the vacuum state of the two initial singly wound strings, so it is hard to disentangle the effect of this vacuum change from any changes due to interactions among the excitations on the string. In the present paper we looked at a 1-loop effect where we have two insertions of $O$. The initial state has two singly wound strings. The first twist converts this to a doubly wound string, and the second twist returns us to two singly wound strings. Thus the vacuum structure of the initial and final states is the same, and any changes to the state can be seen as actual excitations that have been generated on the strings. 

In the single twist case we learned that the effect of a twist operator  generated a squeezed state, described by Bogoliubov coefficients $\gamma^B, \gamma^F$. We have a similar structure for the case of two twists. We have in this paper focused on the effect of the twists $\sigma_2^+$, and left out the action of the supercharge contours that surround these twist insertions. Though these supercharge insertions are essential to the form of the final state after the deformation, they can be split off from the twist insertions and written as supercharge actions on the initial and final states. This computation is straightforward but messy, so we hope to return to it elsewhere and in this paper have  focused only on the more essential physics: the computation of the Bogoliubov coefficients arising from the twists. 

In the single twist case one could find closed form expressions for the Bogoliubov coefficients. In the two twist case the best we could do was write the Bogoliubov coefficients as a finite sum of hypergeometric functions. This form does not make the properties of the coefficients obvious. In particular, we expect the coefficients $\gamma^B_{mn}, \gamma^F_{mn}$. to fall at large values of $m,n$. This is not obvious from any term in the sums. But numerical work showed that the coefficients do fall off as desired. Further, there was a very simple relation between the coefficients with one twist and the coefficients with two twists that can be obtained as a numerical approximation in the limit of large $m,n$. The two twist case also brings in a new feature that was not present in the 1-twist case: the coefficients depend on the separation $\Delta w$ between the twists. We found that there was a simple oscillatory behavior of the coefficients with $\Delta w$.

To obtain a complete analysis of the deformed theory at second order, we must use the Bogoluibov coefficients obtained here, but we must also allow excitations above the vacuum in the initial state, and  apply the  supercharge contours arising from the deformation operators. We expect to return to this in a future work.

\section*{Acknowledgements}
This work is supported in part by DOE grant de-sc0011726.

\appendix
\section{CFT notation and conventions} \label{ap:CFT-notation}

We follow the notation of \cite{acm1, acm2}, which we record here for convenience.
We have 4 real left moving fermions $\psi_1, \psi_2, \psi_3, \psi_4$ which we group into doublets $\psi^{\alpha A}$ as follows:
\be
\begin{pmatrix}
\psi^{++} \cr \psi^{-+}
\end{pmatrix}
=\sqi
\begin{pmatrix}
\psi_1+i\psi_2 \cr \psi_3+i\psi_4
\end{pmatrix}
\ee
\be
\begin{pmatrix}
\psi^{+-} \cr \psi^{--}
\end{pmatrix}
=\sqi
\begin{pmatrix}
\psi_3-i\psi_4 \cr -(\psi_1-i\psi_2)
\end{pmatrix}.
\ee
Here $\alpha=(+,-)$ is an index of the subgroup $SU(2)_L$ of rotations on $S^3$ and $A=(+,-)$ is an index of the subgroup $SU(2)_1$ from rotations in $T^4$. The reality conditions on the individual fermions are
\bea
(\psi_i)^{\dagger} = \psi_i \qquad \Rightarrow \qquad (\psi^{\a A})^{\dagger} = - \epsilon_{\alpha\beta}\epsilon_{AB} \psi^{\beta B} \,.
\eea 
One can introduce doublets $\psi^\dagger$, whose components are given by
\bea
(\psi^\dagger)_{\alpha A} &=& (\psi^{\a A})^{\dagger},
\eea 
from which the reality condition is given by
\be
 (\psi^\dagger)_{\alpha A}=-\epsilon_{\alpha\beta}\epsilon_{AB} \psi^{\beta B}.
\ee
The 2-point functions are
\be
<\psi^{\alpha A}(z)(\psi^\dagger)_{\beta B}(w)>=\delta^\alpha_\beta\delta^A_B{1\over z-w}, ~~~
<\psi^{\alpha A}(z)\psi^{\beta B}(w)>=-\epsilon^{\alpha\beta}\epsilon^{AB}{1\over z-w},
\ee
where we have:
\be
\epsilon_{12}=1, ~~~\epsilon^{12}=-1, ~~~
\psi_A=\epsilon_{AB}\psi^B, ~~~
\psi^A=\epsilon^{AB}\psi_B \,.
\ee
There are 4 real left moving bosons $X_1, X_2, X_3, X_4$, which can be grouped into a matrix:
\be
X_{A\dot A}= \sqi X_i \sigma_i
=\sqi
\begin{pmatrix}
X_3+iX_4 & X_1-iX_2 \\ X_1+iX_2&-X_3+iX_4
\end{pmatrix},
\ee
where $\sigma_i=(\sigma_a, iI)$. The reality condition on the individual bosons is given by
\bea
(X_i)^{\dagger} = X_i \qquad \Rightarrow \qquad (X_{A\dot A})^{\dagger} = - \e^{AB}\e^{\dot A \dot B} X_{B \dot B} \,.
\eea 
One can introduce a matrix, $X^\dagger$, with components  
\be
(X^\dagger)^{A\dot A}~=~ (X_{A\dot A})^{\dagger}~=~\sqi
\begin{pmatrix}
X_3-iX_4& X_1+iX_2\\
X_1-iX_2&-X_3-iX_4
\end{pmatrix},
\ee
from which the reality condition is given by
\bea
(X^\dagger)^{A\dot A}~=~ - \e^{AB}\e^{\dot A \dot B} X_{B \dot B} \,.
\eea 
The 2-point functions are
\be
<\partial X_{A\dot A}(z) (\partial X^\dagger)^{B\dot B}(w)>=-{1\over (z-w)^2}\delta^B_A\delta^{\dot B}_{\dot A}, ~~~
<\partial X_{A\dot A}(z) \partial X_{B\dot B}(w)>={1\over (z-w)^2}\epsilon_{AB}\epsilon_{\dot A\dot B} \,.
\ee

The chiral algebra is generated by the operators
\be
J^a=-{1\over 4}(\psi^\dagger)_{\alpha A} (\sigma^{Ta})^\alpha{}_\beta \psi^{\beta A}
\ee
\be
G^\alpha_{\dot A}= \psi^{\alpha A} \partial X_{A\dot A}, ~~~(G^\dagger)_{\alpha}^{\dot A}=(\psi^\dagger)_{\alpha A} \partial (X^\dagger)^{A\dot A}
\ee
\be
T=-{1\over 2} (\partial X^\dagger)^{A\dot A}\partial X_{A\dot A}-{1\over 2} (\psi^\dagger)_{\alpha A} \partial \psi^{\alpha A}
\ee
\be
(G^\dagger)_{\alpha}^{\dot A}=-\epsilon_{\alpha\beta} \epsilon^{\dot A\dot B}G^\beta_{\dot B}, ~~~~G^{\alpha}_{\dot A}=-\epsilon^{\alpha\beta} \epsilon_{\dot A\dot B}(G^\dagger)_\beta^{\dot B} \,.
\ee
These operators generate the OPE algebra
\be
J^a(z) J^b(z')\sim \delta^{ab} {\h\over (z-z')^2}+i\epsilon^{abc} {J^c\over z-z'}
\ee
\be
J^a(z) G^\alpha_{\dot A} (z')\sim {1\over (z-z')}\h (\sigma^{aT})^\alpha{}_\beta G^\beta_{\dot A}
\ee
\be
G^\alpha_{\dot A}(z) (G^\dagger)^{\dot B}_\beta(z')\sim -{2\over (z-z')^3}\delta^\alpha_\beta \delta^{\dot B}_{\dot A}- \delta^{\dot B}_{\dot A}  (\sigma^{Ta})^\alpha{}_\beta [{2J^a\over (z-z')^2}+{\partial J^a\over (z-z')}]
-{1\over (z-z')}\delta^\alpha_\beta \delta^{\dot B}_{\dot A}T
\ee
\be
T(z)T(z')\sim {3\over (z-z')^4}+{2T\over (z-z')^2}+{\partial T\over (z-z')}
\ee
\be
T(z) J^a(z')\sim {J^a\over (z-z')^2}+{\partial J^a\over (z-z')} 
\ee
\be
T(z) G^\alpha_{\dot A}(z')\sim {{3\over 2}G^\alpha_{\dot A}\over (z-z')^2}  + {\partial G^\alpha_{\dot A}\over (z-z')} \,.
\ee

Note that
\be
J^a(z) \psi^{\gamma C}(z')\sim {1\over 2} {1\over z-z'} (\sigma^{aT})^\gamma{}_\beta \psi^{\beta C} \,.
\ee

The above OPE algebra gives the commutation relations
\begin{eqnarray}
\com{J^a_m}{J^b_n} &=& \frac{m}{2}\delta^{ab}\delta_{m+n,0} + i{\epsilon^{ab}}_c J^c_{m+n}
            \\
\com{J^a_m}{G^\alpha_{\dot{A},n}} &=& \frac{1}{2}{(\sigma^{aT})^\alpha}_\beta G^\beta_{\dot{A},m+n}
             \\
\ac{G^\alpha_{\dot{A},m}}{G^\beta_{\dot{B},n}} &=& \hspace*{-4pt}\epsilon_{\dot{A}\dot{B}}\bigg[
   (m^2 - \frac{1}{4})\epsilon^{\alpha\beta}\delta_{m+n,0}
  + (m-n){(\sigma^{aT})^\alpha}_\gamma\epsilon^{\gamma\beta}J^a_{m+n}
  + \epsilon^{\alpha\beta} L_{m+n}\bigg]\quad\\
\com{L_m}{L_n} &=& \frac{m(m^2-\frac{1}{4})}{2}\delta_{m+n,0} + (m-n)L_{m+n}\\
\com{L_m}{J^a_n} &=& -n J^a_{m+n}\\
\com{L_m}{G^\alpha_{\dot{A},n}} &=& \left(\frac{m}{2}-n\right)G^\alpha_{\dot{A},m+n} \,.
\end{eqnarray}

\section{Ramond vacua notation}\label{RVN}
Here we define our notation for the various Ramond vacua in the untwisted sector.  There are two copies, which are not technically separate Hilbert spaces.  We start with the vacuum
\be
\rmutvket \equiv |v\rangle
\ee
and act on it with various fermion zero modes to construct the other Ramond vacua.  In order to be consistent with ACM, we also require something along the lines of
\be
\rptvket^{(i)} = d_0^{(i)++}d_0^{(i)+-}\rmtvket^{(i)},
\ee
though we do not actually have states containing only one of the two copies.

We now present a table defining our notation for the various vacua.
\bea
|v\rangle & = & \rmutvket \nn
d_0^{(1)+-}|v\rangle &=& |0_R\rangle^{(1)} \otimes |0_R^-\rangle^{(2)}\nn
d_0^{(1)++}|v\rangle &=& |\tilde{0}_R\rangle^{(1)} \otimes |0_R^-\rangle^{(2)}\nn
d_0^{(1)++}d_0^{(1)+-}|v\rangle &=& |0_R^+\rangle^{(1)} \otimes |0_R^-\rangle^{(2)}
\eea
\bea
d_0^{(2)+-}|v\rangle &=& |0_R^-\rangle^{(1)} \otimes |0_R\rangle^{(2)}\nn
d_0^{(1)+-}d_0^{(2)+-}|v\rangle &=& |0_R\rangle^{(1)} \otimes |0_R\rangle^{(2)}\nn
d_0^{(1)++}d_0^{(2)+-}|v\rangle &=& |\tilde{0}_R\rangle^{(1)} \otimes |0_R\rangle^{(2)}\nn
d_0^{(1)++}d_0^{(1)+-}d_0^{(2)+-}|v\rangle &=& |0_R^+\rangle^{(1)} \otimes |0_R\rangle^{(2)}
\eea
\bea
d_0^{(2)++}|v\rangle &=& |0_R^-\rangle^{(1)} \otimes |\tilde{0}_R\rangle^{(2)}\nn
d_0^{(1)+-}d_0^{(2)++}|v\rangle &=& |0_R\rangle^{(1)} \otimes |\tilde{0}_R\rangle^{(2)}\nn
d_0^{(1)++}d_0^{(2)++}|v\rangle &=& |\tilde{0}_R\rangle^{(1)} \otimes |\tilde{0}_R\rangle^{(2)}\nn
d_0^{(1)++}d_0^{(1)+-}d_0^{(2)++}|v\rangle &=& |0_R^+\rangle^{(1)} \otimes |\tilde{0}_R\rangle^{(2)}
\eea
\bea
d_0^{(2)++}d_0^{(2)+-}|v\rangle &=& |0_R^-\rangle^{(1)} \otimes |0_R^+\rangle^{(2)}\nn
d_0^{(1)+-}d_0^{(2)++}d_0^{(2)+-}|v\rangle &=& |0_R\rangle^{(1)} \otimes |0_R^+\rangle^{(2)}\nn
d_0^{(1)++}d_0^{(2)++}d_0^{(2)+-}|v\rangle &=& |\tilde{0}_R\rangle^{(1)} \otimes |0_R^+\rangle^{(2)}\nn
d_0^{(1)++}d_0^{(1)+-}d_0^{(2)++}d_0^{(2)+-}|v\rangle &=& |0_R^+\rangle^{(1)} \otimes |0_R^+\rangle^{(2)}.
\eea
These relations hold for both the initial (pre-twists) and final (post-twists) sectors.

\section{Proof of the exponential form}\label{GeneralFormAppendix}
In this section we prove the exponential form of $|\chi(w_1,w_2)\rangle$ for the case of fermions in the NS sector.  The other cases may be treated similarly.

The method of the proof is straightforward.  We look at capping the state $|\chi(w_1,w_2)\rangle$ with a general femionic NS state $\hat{Q}^{\dagger} \nsnsket$.  This provides the relationship
\bea
\mathcal{A}\left ( \hat{Q} \right ) &\equiv&{\nsnsbra \hat{Q} |\chi(w_1,w_2)\rangle \over \nsnsbra \chi(w_1,w_2)\rangle} ~ = ~ {\nstbra \hat{Q}' \nstket \over \nstbra \nstclose}.\label{GeneralMappingRelation}
\eea
The rightmost expression can be calculated from the behaior of the fermion modes under coordinate transformations and spectral flows, while the middle expression can be calculated from our guess for the form of $|\chi(w_1,w_2)\rangle$.  Showing that these two methods are consistent demonstrates that we have the correct form for $|\chi(w_1,w_2)\rangle$.

\subsection{Middle expression}
Let us begin by taking a close look at the form presented in (\ref{generalform}).  Because we are working with pairs of fermion creation operators in the exponent, all of the terms in the exponent commute.  We can thus rewrite the fermion portion in the NS sector as
\bea
|\chi(w_1,w_2)\rangle^{F}_{NS} & = & e^{\sum\limits_{(i),(j)}\sum\limits_{r,s > 0}\g^{F(i)(j)}_{NS,rs} \left ( d^{(i)f,++}_{-r}d^{(j)f,--}_{-s} - d^{(i)f,+-}_{-r}d^{(j)f,-+}_{-s} \right )}\nsnsket\nn
&=& \prod_{(i),(j)}\prod_{r,s>0}e^{\g^{F(i)(j)}_{NS,rs} \left ( d^{(i)f,++}_{-r}d^{(j)f,--}_{-s} - d^{(i)f,+-}_{-r}d^{(j)f,-+}_{-s} \right )}\nsnsket \nn
&=& \prod_{A,B}\prod_{(i),(j)}\prod_{r,s>0}\left [ 1 + \e_{AB}\g^{F(i)(j)}_{NS,rs}  d^{(i)f,+A}_{-r}d^{(j)f,-B}_{-s} \right ]\nsnsket.\qquad
\eea

From this form, we can see clearly that any state with an odd number of fermion excitations will have no overlap with $|\chi(w_1,w_2)\rangle$.  So let us start with a two-excitation state.  This case is quite simple:
\bea
\mathcal{A}\left(d^{(i)f,+A}_r d^{(j)f,-B}_s\right) &=& {\nsnsbra d^{(i)f,+A}_r d^{(j)f,-B}_s |\chi(w_1,w_2)\rangle \over \nsnsbra \chi(w_1,w_2)\rangle} \nn
&=& \e^{AC}\e^{BD}\e_{CD}\g^{F(i)(j)}_{NS,rs}\nn
&=& \e^{AB}\g^{F(i)(j)}_{NS,rs},\label{2PointAmplitude}
\eea
where the fact that the coefficient is proportional to $\e^{AB}$ simply means that the combinations $d^{++}d^{--}$ and $d^{+-}d^{-+}$ yield coefficients which differ only by an overall sign.

Now what happens when we cap with a state that contains more fermion excitations?  We know that we must have an even number of excitations, so they come in pairs.  We can thus calculate $\mathcal{A}$ in two steps.  First, we write out all possible ways to group the fermion excitations into pairs, accounting for the overal sign required to anticommute the operators into the appropriate pairings.  Then each pairing combination provides a contribution equal to the product of the amplitude for each individual pair within that combination.  Adding the contributions for each combination then gives us the total amplitude.  We can write this schematically as:
\bea
\mathcal{A}\left ( d_1 d_2 d_3 \ldots \right ) &=& \sum_{\text{pairing combinations}} (-1)^p\left (\prod_{\{d_i d_j \}}\mathcal{A}\left ( d_i d_j \right ) \right ), \label{4PointAmplitude}
\eea
where $p$ is the number of anticommutations we need to perform to achieve the pairing configuration.

\subsection{Right expression}
Let us compare the previous result to the calculation in the $t$ plane.  In Section \ref{Outline} we outlined a series of coordinate maps and spectral flows which bring $|\chi(w_1,w_2)\rangle$ to an empty NS vacuum in the $t$ plane.  Under any combination of such transformations, the modes $d^{(i)f,\a A}_r$ behave in general as:
\bea
d^{(1)f,\a A}_r & \to & d'^{(1)f,\a A}_r ~=~ {1\over 2\pi i}\oint\limits_{t=\infty} \psi^{\a A}(t) h^{\a}_r(t) \diff t\\
d^{(2)f,\a A}_r & \to & d'^{(2)f,\a A}_r ~=~  {1\over 2\pi i}\oint\limits_{t=0} \psi^{\a A}(t) h^{\a}_r(t) \diff t,
\eea
where the function $h_r^{\a}$ is identical for each copy but will in general depend on the original mode number $r$ and the relevant spectral flow charge $\a$.

Now we expand the function $h_r^{\a}$ as a polynomial in $t$, which can in general include negative powers of $t$.  It turns out that the function $h_r^{\a}$ consists of a product of powers of $t$ and binomials in $t$, which makes such an expansion straightforward.  We perform the expansion with large $t$ for Copy 1 and small $t$ for copy 2.  We then find:
\bea
d'^{(1)f,\a A}_r & = & \sum_{p=-\infty}^{\infty}C^{(1)\a}_{rp}{1\over 2\pi i}\oint\limits_{t=\infty} \psi^{\a A}(t) t^p \diff t\\
d'^{(2)f,\a A}_r & = & \sum_{p=-\infty}^{\infty}C^{(2)\a}_{rp}{1\over 2\pi i}\oint\limits_{t=0} \psi^{\a A}(t) t^p \diff t,
\eea
where the coefficients are not in general equal because we are expanding the function in two different regions.

Recall that after these coordinate maps and spectral flows, we have removed all insertions in the $t$ plane.  We can thus close all punctures with the NS vacuum and smoothly deform our contours.  This allows us to deform the contour around $t = \infty$ for the copy 1 mode into a contour around $t=0$.  We then find that in general:
\bea
d'^{(i)f,\a A}_r & = & \sum_{p=-\infty}^{\infty}C^{(i)\a}_{rp}{1\over 2\pi i}\oint\limits_{t=0} \psi^{\a A}(t) t^p \diff t\\
&=& \sum_{p=-\infty}^{\infty} C^{(i)\a}_{rp} \tilde{d}^{\a A}_p. \label{GeneralFermionExpansion}
\eea

From this expression we see that when capping with a two-excitation state, the right side of (\ref{GeneralMappingRelation}) becomes:
\bea\label{A2tPlane}
\mathcal{A}\left(d^{(i)f,+A}_r d^{(j)f,-B}_s\right) &=& \sum_{p,q=-\infty}^{\infty}C^{(i)+}_{rp}C^{(j)-}_{sq}{\nstbra : \tilde{d}^{+A}_p \tilde{d}^{-B}_q :\nstket \over \nstbra\nstclose},
\eea
where the colons indicate radial ordering of the contours in the modes, which may not be the same as the initial $\t$ ordering on the cylinder, as copy 1 modes always map to contours at larger radii than copy 2 modes.

The amplitude in the numerator on the right of (\ref{A2tPlane}) is only nonzero when the two fermion modes can contract together, with the outer mode annihilating the inner mode.  However, we know the anticommutator is:
\bea
\left \{ \tilde{d}^{+A}_p , \tilde{d}^{-B}_q \right \} &=& -\e^{+-}\e^{AB} \d_{p+q,0},
\eea
which is proportional to $\e^{AB}$.  We thus find that for any fermion pair,
\bea
\mathcal{A}\left(d^{(i)f,+A}_r d^{(j)f,-B}_s\right) &=& \e^{AB}D^{(i)(j)}_{rs},
\eea
which matches the result from (\ref{2PointAmplitude}).

Now what happens when we cap with a state that contains more fermion excitations?  In general, we have:
\bea
\mathcal{A}\left( d_1 d_2 d_3 \ldots \right) &=& \sum_{p_1,p_2,p_3,\ldots} \left (C_{p_1} C_{p_2} C_{p_3} \ldots \right ) {\nstbra : \tilde{d}_{p_1} \tilde{d}_{p_2}\tilde{d}_{p_3}\ldots :\nstket \over \nstbra\nstclose}.
\eea
It is clear that the right side vanishes when there are an odd number of fermion modes, as such a case will always leave at least one uncontracted mode to annihilate the vacuum bra or ket.  With an even number of modes, the amplitude can be calculated by performing all possible combinations of contractions in the expression on the right side and adding each contribution together.  However, we also have the relation
\bea
{\nstbra \contraction{}{\tilde{d}}{{}_{p_1}}{\tilde{d}} \tilde{d}_{p_1}\tilde{d}_{p_2}\contraction{}{\tilde{d}}{{}_{p_3}}{\tilde{d}} \tilde{d}_{p_3}\tilde{d}_{p_4}\nstket\over \nstbra\nstclose} &=& {\nstbra :\tilde{d}_{p_1}\tilde{d}_{p_2}:\nstket \over \nstbra\nstclose} {\nstbra:\tilde{d}_{p_3}\tilde{d}_{p_4}:\nstket\over \nstbra\nstclose}.\qquad
\eea
We thus find:
\bea
\mathcal{A}\left (d_1 d_2 d_3 \ldots \right )&=& \sum_{\text{pairing combinations}} (-1)^p\left [ \prod_{\{p_i,p_j\}} \left (\sum_{p_i,p_j}C_{p_i}C_{p_j}{\nstbra : \tilde{d}_{p_i} \tilde{d}_{p_j} :\nstket \over \nstbra\nstclose}\right )\right ] \nn
&=& \sum_{\text{pairing combinations}} (-1)^p\left (\prod_{\{d_i d_j \}}\mathcal{A}\left ( d_i d_j \right ) \right ), \label{4PointTAmplitude}
\eea
where $p$ is again the number of anticommutations we must perform to reach the particular pair combination.  Since this overall sign depends only on how we group the operators in $\hat{Q}$ and not on any of the specifics of the coordinate maps or spectral flows, each $p$ that appears here takes the same value as in in (\ref{4PointAmplitude}).  Thus the two relations are identical, and we see that the claimed form for $|\chi(w_1,w_2)\rangle$ is correct.

\section{Proof of copy 1-copy 2 interchange symmetry}\label{CopySymmetry}
Using the vacua listed in Appendix (\ref{RVN}), we can look at the effects of $\g^{F(i)(j)}_{R+-,mn}$ and check the desired $(1) \leftrightarrow (2)$ symmetry.  We first note that $\g^{F(2)(1)}_{R+-;0,0}=1$.  We will also introduce the shorthand
\bea
\g^{F(2)(1)}_{R+-;m,0} ~\equiv~ A_m \implies \g^{F(1)(1)}_{R+-;m,0}~=~ -A_m.
\eea
With this notation, we find:
\bea
|\chi(w_1,w_1)\rangle &=& \left (1+d^{(2)f,++}_{0}d^{(1)f,--}_{0}\right)\left(1-d^{(2)f,+-}_{0}d^{(1)f,-+}_{0}\right)\left(1+A_md^{(2)f,++}_{-m}d^{(1)f,--}_0\right)\nn
&&{}\times\left(1-A_{m'}d^{(2)f,+-}_{-m'}d^{(1)f,-+}_0\right)\left(1-A_nd^{(1)f,++}_{-n}d^{(1)f,--}_0\right)\nn
&&{}\times\left(1+A_{n'}d^{(1)f,+-}_{-n'}d^{(1)f,-+}_0\right)d_0^{(1)f,++}d_0^{(1)f,+-}e^{\hat{Q}}|0_R^-\rangle^{(1)}|0_R^-\rangle^{(2)},\label{ChiZeroes}
\eea
where $\hat{Q}$ now contains no zero modes and the repeated mode indices are summed over positive integers

Noting the relations (\ref{BosonRelation1}), (\ref{BosonRelation2}) and (\ref{symmetryrequirements}), we see that the operator $\hat{Q}$ is symmetric under $(1) \leftrightarrow (2)$.  We must then inqure as to the symmetry of the rest of (\ref{ChiZeroes}).  In doing so, let us write $P$ as the operation which swaps copies $(1)$ and $(2)$.  We then note that the notation outlined in (\ref{RVN}) implies:
\bea
P\left ( |0_R\rangle^{(1)}|\tilde{0}_R\rangle^{(2)} \right ) &=& P \left ( d^{(1)f,+-}_0d^{(2)f,++}_0|0_R^-\rangle^{(1)}|0_R^-\rangle^{(2)}\right ) \nn
&=& d^{(2)f,+-}_0d^{(1)f,++}_0|0_R^-\rangle^{(1)}|0_R^-\rangle^{(2)} \nn
&=& -d^{(1)f,++}_0d^{(2)f,+-}_0|0_R^-\rangle^{(1)}|0_R^-\rangle^{(2)}\nn
&=&-|\tilde{0}_R\rangle^{(1)}|0_R\rangle^{(2)}.
\eea
And since $P^2=1$, we also have:
\bea
P\left ( |\tilde{0}_R\rangle^{(1)}|0_R\rangle^{(2)} \right ) &=& -|0_R\rangle^{(1)}|\tilde{0}_R\rangle^{(2)}.
\eea
On the other hand,
\bea
P\left ( |0_R^+\rangle^{(1)}|0_R^-\rangle^{(2)} \right ) &=& |0_R^-\rangle^{(1)}|0_R^+\rangle^{(2)}\\
P\left(|0_R^-\rangle^{(1)}|0_R^+\rangle^{(2)}\right)&=&|0_R^+\rangle^{(1)}|0_R^-\rangle^{(2)}.
\eea
Using these relations, we find:
\bea
|v\rangle &\equiv& \left(1+d^{(2)f,++}_{0}d^{(1),f--}_{0}\right)\left(1-d^{(2)f,+-}_{0}d^{(1)f,-+}_{0}\right)d_0^{(1)f,++}d_0^{(1)f,+-}|0_R^-\rangle^{(1)}|0_R^-\rangle^{(2)}\nn
&=& |0_R^+\rangle^{(1)}|0_R^-\rangle^{(2)} + |0_R\rangle^{(1)}|\tilde{0}_R\rangle^{(2)} - |\tilde{0}_R\rangle^{(1)}|0_R\rangle^{(2)} + |0_R^-\rangle^{(1)}|0_R^+\rangle^{(2)},
\eea
which is symmetric under $P$.  We now write:
\bea
|\chi(w_1,w_1)\rangle &=& \left(1+A_md^{(2)f,++}_{-m}d^{(1)f,--}_0\right)\left(1-A_{m'}d^{(2)f,+-}_{-m'}d^{(1)f,-+}_0\right)\nn
&&{}\times\left(1-A_nd^{(1)f,++}_{-n}d^{(1)f,--}_0\right)\left(1+A_{n'}d^{(1)f,+-}_{-n'}d^{(1)f,-+}_0\right)e^{\hat{Q}}|v\rangle.\label{Intermediate}
\eea

At this point it is necessary to make use of an interesting property of the state $|v\rangle$.  When applying zero modes to this state, we notice:
\bea
d^{(1)f,--}_0 |v\rangle &=& -|0_R\rangle^{(1)}|0_R^-\rangle^{(2)} + |0_R^-\rangle^{(1)}|0_R\rangle^{(2)}\nn
d^{(2)f,--}_0 |v\rangle &=& |0_R\rangle^{(1)}|0_R^-\rangle^{(2)} - |0_R^-\rangle^{(1)}|0_R\rangle^{(2)}\nn
d^{(1)f,-+}_0 |v\rangle &=&  - |\tilde{0}_R\rangle^{(1)}|0_R^-\rangle^{(2)} + |0_R^-\rangle^{(1)}|\tilde{0}_R\rangle^{(2)}\nn
d^{(2)f,-+}_0 |v\rangle &=&  |\tilde{0}_R\rangle^{(1)}|0_R^-\rangle^{(2)} - |0_R^-\rangle^{(1)}|\tilde{0}_R\rangle^{(2)}.
\eea
Thus whenever we have a copy (1) zero mode with negative charge acting on $|v\rangle$ we can substitute a copy (2) zero mode in its place:
\bea
d^{(1)f,-\a}_0|v\rangle &=& -d^{(2)f,-\a}_0|v\rangle.
\eea
Applying this substitution to the second line of (\ref{Intermediate}), we have:
\bea
|\chi(w_1,w_1)\rangle &=& \left(1+A_md^{(2)f,++}_{-m}d^{(1)f,--}_0\right)\left(1-A_{m'}d^{(2)f,+-}_{-m'}d^{(1)f,-+}_0\right)\nn
&&{}\times\left(1+A_nd^{(1)f,++}_{-n}d^{(2)f,--}_0\right)\left(1-A_{n'}d^{(1)f,+-}_{-n'}d^{(2)f,-+}_0\right)e^{\hat{Q}}|v\rangle,\qquad
\eea
which is manifestly symmetric.

\section{Proof of $\g^{F}$ and $\g^{B}$ Relations}\label{GammaRelations}
In this section we analytically prove the relations:
\bea
\g^{B(1)(1)}_{mn}&=&-\g^{B(2)(1)}_{mn}\cr\cr
\left [ \g^{F(1)(1)}_{R+-,mn}\right ]_{n>0}&=&-\left[\g^{F(2)(1)}_{R+-,mn}\right]_{m,n>0}.
\eea

\subsection{Proof of $\g^{B(1)(1)}_{mn}=-\g^{B(2)(1)}_{mn}$}
Beginning with $\g^{B(i)(j)}_{mn}$, we have the relation
\bea
\g^{B(1)(1)}_{mn}=-\g^{B(2)(1)}_{mn}
\eea
for $a\neq 0$. 
Let us write the expressions for both functions:
\bea
\g^{B(1)(1)}_{mn}&=&\frac{1}{mn}\sum_{l=1}^{n}l{}^{n}C_{l}{}^{m}C_{l}a^{l}b^{m+n-l}{}_2F_1\left(-n,l-n;l+1;\frac{a}{b}\right){}_2F_1\left(-m,l-m;l+1;\frac{a}{b}\right),\nn
\eea
and
\bea
\g^{B(2)(1)}_{mn}&=&-\frac{1}{mn}\sum_{l=1}^{n}l{}^{m}C_{l}{}^{n}C_{l}a^{m}b^{n}{}_2F_1\left(-m,l-m;l+1;\frac{b}{a}\right){}_2F_1\left(-n,l-n;l+1;\frac{a}{b}\right).\nn
\label{gammab12}
\eea
Let us investigate $\g^{B(2)(1)}$. Considering the \emph{first} hypergeometric function with the identity:
\bea
{}_{2}F_{1}(a,b,c;z)&=&\frac{\G(1-a)\G(c)}{\G(b)\G(c-a-b+1)}\left(-1\over z\right)^{a}\left(1-\frac{1}{z}\right)^{c-a-b}\nn
&&~\times~ {}_{2}F_{1}\left(1-b,c-b;c-a-b+1;1-\frac{1}{z}\right)\nn
&& +~ {\G(1-a)\G(c)\over \G(b-a+1)\G(c-b)}\left({1\over z}\right)^{b}{}_{2}F_{1}(b,b-c+1;b-a+1;{1\over z}),\qquad
\label{identity}
\eea
we find:
\bea
{}_2F_1\left(-m,l-m;l+1;\frac{b}{a}\right)&=&\frac{\G(m+1)\G(l+\h)}{\G(l-m)\G(2m+2)}\left(-a\over b\right)^{-m}\left(1-\frac{a}{b}\right)^{2m+1}\nn
&&~\times~{}_{2}F_{1}\left(m-l+1,m+1;2m+2;1-\frac{a}{b}\right)\nn
&& +~ {\G(m+1)\G(l+1)\over \G(l+1)\G(m+1)}\left({a\over b}\right)^{l-m}{}_{2}F_{1}(l-m,-m;l+1;{a\over b}).\nn
\eea
The first term vanishes because 
\bea
\frac{1}{\G(l-m)}=0~~ \text{when} ~~  l-m \leq 0,
\eea
which is always the case. We also note that the ratio of gamma functions for the second term is one. Using the symmetry relation
\bea
{}_{2}F_{1}(a,b,c,z)={}_{2}F_{1}(b,a,c,z),
\label{interchange symmetry}
\eea
we find that
\bea
{}_{2}F_{1}(l-m,-m;l+1;{a\over b})={}_{2}F_{1}(-m,l-m;l+1;{a\over b}).
\eea
Inserting this into (\ref{gammab12}),  we obtain:
\bea
\g^{B(2)(1)}_{mn}&=&-\frac{1}{mn}\sum_{l=1}^{n}l{}^{m}C_{l}{}^{n}C_{l}a^{l}b^{m+n-l}{}_2F_1\left(-m,l-m;l+1;\frac{a}{b}\right){}_2F_1\left(-n,l-n;l+1;\frac{a}{b}\right)\cr\cr
&=&-\g^{B(1)(1)}_{mn}.
\eea

\subsection{Proof of $\left [ \g^{F(1)(1)}_{R+-,mn}\right ]_{n>0}=-\left[\g^{F(2)(1)}_{R+-,mn}\right]_{m,n>0}$}
Here we prove the $\g^{F(i)(j)}_{R+-,mn}$ relation
\bea
\left [ \g^{F(1)(1)}_{R+-,mn}\right ]_{n>0}=-\left[\g^{F(2)(1)}_{R+-,mn}\right]_{m,n>0}
\eea
for $a\neq 0$.  Let us first write expressions for both functions. We have:
\bea
\left[\g^{F(1)(1)}_{R+-,mn}\right]_{n>0}&=& -\sum_{p = {3\over 2}}^{n+\h}{}^{n-1}C_{n-p+\h}{}^m C_{p-\h}a^{p-\h}b^{m+n-p+\h}\nn
&&\quad{}\times {}_2 F_1 \left ( 1-n, p-n-\h; p-\h;{a\over b}\right ) {}_2 F_1 \left ( p-m-\h,-m;p+\h;{a\over b}\right ) \nn
&&{}+\sum_{p = {3\over 2}}^{n-{3\over2}}{}^{n-1}C_{n-p-{3\over2}}{}^m C_{p-\h}a^{p+\h}b^{m+n-p-\h}\nn
&&\quad{}\times {}_2 F_1 \left ( 1-n, p-n+{3\over2}; p+{3\over2};{a\over b}\right ) {}_2 F_1 \left ( p-m-\h,-m;p+\h;{a\over b}\right ), \nn
\eea
and 
\bea
\left[\g^{F(2)(1)}_{R+-,mn}\right]_{m,n>0}&=& \sum_{p={1\over 2}}^{\min\left(m-\h,n-\h\right)}{}^{n-1}C_{n-p-\h}{}^m C_{m-p-\h}a^m b^n\nn
&&{}\times {}_2 F_1 \left(1-n,p-n+\h;p+\h;{a\over b}\right){}_2 F_1 \left(p-m+\h,-m;p+\frac{3}{2};{b\over a}\right) \nn
&&{}-\sum_{p={1\over2}}^{\min\left(m-\h,n-{5\over2}\right)}{}^{n-1}C_{n-p-{5\over2}}{}^m C_{m-p-\h}a^{m+1}b^{n-1}\nn
&&{}\times {}_2 F_1 \left(1-n,p-n+{5\over2};p+{5\over2};{a\over b}\right){}_2 F_1 \left(p-m+\h,-m;p+{3\over2};{b\over a}\right).\nn
\eea
We shall now investigate $\g^{F(2)(1)}$. First let us make the index shift
\bea
p\to p-1.
\eea
For $\g^{F(2)(1)}$ this gives
\bea
\left[\g^{F(2)(1)}_{R+-,mn}\right]_{m,n>0}&=& \sum_{p={3\over 2}}^{\min\left(m+\h,n+\h\right)}{}^{n-1}C_{n-p+\h}{}^m C_{m-p+\h}a^m b^n\nn
&&{}\times {}_2 F_1 \left(1-n,p-n-\h;p-\h;{a\over b}\right){}_2 F_1 \left(p-m-\h,-m;p+\frac{1}{2};{b\over a}\right) \nn
&&{}-\sum_{p={3\over2}}^{\min\left(m+\h,n-{3\over2}\right)}{}^{n-1}C_{n-p-{3\over2}}{}^m C_{m-p+\h}a^{m+1}b^{n-1}\nn
&&{}\times {}_2 F_1 \left(1-n,p-n+{3\over2};p+{3\over2};{a\over b}\right){}_2 F_1 \left(p-m-\h,-m;p+{1\over2};{b\over a}\right).\nn
\label{gamma21}
\eea
Now using the identity given in (\ref{identity}) to transform the second hypergeometric function of each term as well as the interchange symmetry given in (\ref{interchange symmetry}), we obtain:
\bea
{}_{2}F_{1}\left(p-m-\h,-m,p-\h;{b\over a}\right)&=&{}_{2}F_{1}\left(-m,p-m-\h,p-\h;{b\over a}\right)\cr\cr
&=&\frac{\G(m+1)\G(p+1/2)}{\G(p-m-1/2)\G(2m+2)}\left(-a\over b\right)^{-m}\left(1-\frac{a}{b}\right)^{2m+1}\cr\cr
&&\quad\times{}_{2}F_{1}\left(m-p+3/2,m;2m+2;1-\frac{a}{b}\right)\cr\cr
&&\quad +~ \frac{\G(m+1)\G(p+1/2)}{\G(m+1)\G(p+1/2)}\cr\cr
&&\quad~~~\left({a\over b}\right)^{p-m-1/2}{}_{2}F_{1}\left(p-m-{1\over 2},-m;p+{1\over 2};{a\over b}\right).
\nn
\label{identity}
\eea
We see that the first term vanishes because
\bea
\frac{1}{\G\left(p-m-1/2\right)}=0~~\text{when}~~p\leq m+1/2,
\eea
which is always the case.  Additionally, the ratio of gamma functions in the second term is one. Therefore we have
\bea
{}_{2}F_{1}\left(p-m-\h,-m,p-\h;{b\over a}\right)=\left({a\over b}\right)^{p-m-1/2}{}_{2}F_{1}\left(p-m-{1\over 2},-m;p+{1\over 2};{a\over b}\right).
\eea
Inserting this term into (\ref{gamma21}), we obtain:
\bea
\left[\g^{F(2)(1)}_{R+-,mn}\right]_{m,n>0}&=& \sum_{p={3\over 2}}^{\min\left(m+\h,n+\h\right)}{}^{n-1}C_{n-p+\h}{}^m C_{m-p+\h}a^{p-{1\over 2}} b^{m+n-p+\h}\nn
&&{}\times {}_2 F_1 \left(1-n,p-n-\h;p-\h;{a\over b}\right){}_2 F_1 \left(p-m-\h,-m;p+\frac{1}{2};{a\over b}\right) \nn
&&{}-\sum_{p={3\over2}}^{\min\left(m+\h,n-{3\over2}\right)}{}^{n-1}C_{n-p-{3\over2}}{}^m C_{m-p+\h}a^{p+\h} b^{m+n-p-\h}\nn
&&{}\times {}_2 F_1 \left(1-n,p-n+{3\over 2};p+{3\over 2};{a\over b}\right){}_2 F_1 \left(p-m-\h,-m;p+{1\over 2};{a\over b}\right)\crcr
&=&-\left[\g^{F(1)(1)}_{R+-,mn}\right]_{n>0}.
\eea

\section{$\g^{B(i)(j)}_{mn}$ and $\g^{F(i)(j)}_{R+-,mn}$ relation}\label{boson fermion relation}
Here we prove the relationship 
\bea
\g^{F(i)(j)}_{R+-,mn}=-m\g^{B(i)(j)}_{mn}.
\eea
We start by applying $G^{+(p)}_{0,+}$ to $|\chi(w_{1},w_{2})\rangle$.
\bea
\sum_{(p)}G^{(p)+}_{0,+}|\chi(w_{1},w_{2})\rangle&=&\sum_{(p)}G^{(p)+}_{0,+}\s_{2}^{+}(w_{2})\s_{2}^{+}(w_{1})|0_{R}^{-}\rangle^{(1)}|0_{R}^{-}\rangle^{(2)}\nn
&=&\left({1\over 2\pi i}\int_{w_{2}}dwG^{+}_{+}(w)\s_{2}^{+}(w_{2})\right)\s_{2}^{+}(w_{1})|0_{R}^{-}\rangle^{(1)}|0_{R}^{-}\rangle^{(2)}\nn
&&~ + ~ \s_{2}^{+}(w_{2})\left({1\over 2\pi i}\int_{w_{1}}dwG^{+}_{+}(w)\s_{2}^{+}(w_{1})\right)|0_{R}^{-}\rangle^{(1)}|0_{R}^{-}\rangle^{(2)}\cr\cr
&&~ + ~ \s_{2}(w_{1})\s_{2}^{+}(w_{1})\sum_{(j)}G^{(j)+}_{0,+}|0_{R}^{-}\rangle^{(1)}|0_{R}^{-}\rangle^{(2)}\nn
&=& 0.
\label{supercharge chi relation}
\eea
where $j,p\in\lbrace 1,2\rbrace$. We have used the fact that $G^{+}_{0,\dot{A}}$ acting on a single $\s_{2}^{+}(w_{i})$ is \textit{zero}\footnote{This can be seen clearly from mapping to the $t$ plane and performing the appropriate spectral flow, giving an annihilation operator on the $t$ plane vacuum rather than the creation operator that a $G^-_{0,\dot A}$ contour produces.}, as well as the fact that $G^{+}_{0,+}$ annihilates the negative Ramond vacuum. More explicitly:

\bea
&&{1\over 2\pi i}\int_{w_{i}}dwG^{+}_{+}(w)\s_{2}^{+}(w_{i})=0\nn
&&G^{(j)+}_{0,+}|0_{R}^{-}\rangle^{(1)}|0_{R}^{-}\rangle^{(2)}=0.
\eea

Using the form of $|\chi(w_{1},w_{2})\rangle$ given in (\ref{chi state}), where we have excluded the $NS$ sector states, we find:
\bea
0 &=&\sum_{(p)}G^{(p)+}_{0,+}|\chi(w_1,w_2)\rangle\nn
&=& \sum_{(p)}G^{(p)+}_{0,+}C_{R+-}\text{exp}\left [ \sum_{(i),(j)}\sum_{k,l > 0}\g^{B(i)(j)}_{kl} \left ( -\a^{(i)f}_{++,-k}\a^{(j)f}_{--,-l} + \a^{(i)f}_{+-,-k}\a^{(j)f}_{-+,-l} \right ) \right ] \nn
&&\quad{}\times \text{exp}\left [ \sum_{(i),(j)}\sum_{k,l \geq 0}\g^{F(i)(j)}_{R+-,kl} \left ( d^{(i)f,++}_{-k}d^{(j)f,--}_{-l} - d^{(i)f,+-}_{-k}d^{(j)f,-+}_{-l} \right ) \right ] \rpmket \ \nn
&=&C_{R+-}\sum_{(i),(j),(p)}\left[\,\sum_{k,l > 0}\g^{B(i)(j)}_{kl} \left[G^{(p)+}_{0,+},  -\a^{(i)f}_{++,-k}\a^{(j)f}_{--,-l} + \a^{(i)f}_{+-,-k}\a^{(j)f}_{-+,-l} \right]\right.\nn
&&\left.\quad{}+\sum_{k,l \geq 0}\g^{F(i)(j)}_{R+-,kl} \left\{ G^{(p)+}_{0,+},  d^{(i)f,++}_{-k}d^{(j)f,--}_{-l} - d^{(i)f,+-}_{-k}d^{(j)f,-+}_{-l} \right\}\right]|\chi(w_1,w_2)\rangle  \nn
&=&C_{R+-}\sum_{(i),(j),(p)}\left[\,\sum_{k,l > 0}\g^{B(i)(j)}_{kl}\left(-\a^{(i)f}_{++,-k}\left[G^{(p)+}_{0,+},\a^{(j)f}_{--,-l}\right]+\left[G^{(p)+}_{0,+},\a^{(i)f}_{+-,-k}\right]\a^{(j)f}_{-+,-l}\right)\right.\nn
&&\left.\quad{}+\sum_{k,l \geq 0}\g^{B(i)(j)}_{kl}\left( - d^{(i)f,++}_{-k}\lbrace G^{(p)+}_{0,+}, d^{(j)f,--}_{-l}\rbrace + \lbrace G^{(p)+}_{0,+},d^{(j)f,-+}_{-l}\rbrace d^{(i)f,+-}_{-k}  \right)\right]|\chi(w_1,w_2)\rangle.\nn
\eea

We now apply the (anti)commutation relations:
\bea
\left[G^{(i)\a}_{\dot{A},0},\a^{(j)}_{B\dot{B},m}\right]&=&-im\e_{AB}\e_{\dot{A}\dot{B}}\d^{(i)(j)}d^{(i)\a A}_{m}\cr
\lbrace G^{(i)\a}_{\dot{A},0},d^{(j),\b B}_{m} \rbrace&=&i\e^{\a\b}\e^{AB}\d^{(i)(j)}\a^{(i)}_{A\dot{A},m}.
\eea
This gives:

\bea
0 &=& C_{R+-} \sum_{(i),(j),(p)}\left[\,\sum_{k,l > 0}\g^{B(i)(j)}_{kl}\left(  \a^{(i)f}_{++,-k}\left(-il\right)\d^{(p)(j)}d^{(p)f,++}_{-l}+(-ik)\d^{(p)(i)}d^{(p)f,+-}_{-k}\a^{(j)f}_{-+,-l}  \right)\right.\nn
&&\left.\quad{}+\sum_{k,l \geq 0}\g^{F(i)(j)}_{kl}\left(  -d^{(i)f,++}_{-k}(i)\d^{(p)(j)}\a_{++,-l}^{(p)f}+(-i)\d^{(p)(j)}\a_{-+,-l}^{(p)f}d^{(i)f,+-}_{-k}\right)\right]|\chi(w_1,w_2)\rangle \nn
&=&-iC_{R+-}\sum_{(i),(p)}\sum_{k,l>0}\left(l\g^{B(i)(p)}_{kl}\a_{++,-k}^{(i)f}d^{(p)f,++}_{-l}+k\g^{B(p)(i)}_{kl}\a_{-+,-l}^{(i)f}d^{(p)f,+-}_{-k}\right.\nn
&&\qquad\qquad\left.{}+\g_{kl}^{F(i)(p)}\a_{++,-l}^{(p)f}d^{(i)f,++}_{-k}+\g^{F(i)(p)}_{kl}\a_{-+,-l}^{(p)f}d^{(i)f,+-}_{-k}\right)|\chi(w_{1},w_{2})\rangle\nn
&&\quad{}-iC_{R+-}\sum_{(i),(p)}\sum_{l > 0}\left(\g^{F(i)(p)}_{0,l}\a_{++,-l}^{(p)f}d^{(i)f,++}_{0} +\g^{F(i)(p)}_{0,l}\a_{-+,-l}^{(p)f}d_{0}^{(i)f,+-}\right)|\chi(w_{1},w_{2})\rangle,
\label{Gplus1}
\eea
where we have used the fact that all bosonic zero modes annihilate $|\chi(w_1,w_2)\rangle$.  The fermionic zero modes do not annihilate this state, but from (\ref{symmetryrequirements}) we know that
\bea
\left[\g^{F(p)(i)}_{R+-;0,l}\right]_{l>0}=0, \quad i, j\in {1,2}.
\eea
Thus the last line of (\ref{Gplus1}) vanishes, and we are left with
\bea
0&=&-iC_{R+-}\sum_{(i),(p)}\sum_{k,l>0}\left(l\g^{B(i)(p)}_{kl}\a_{++,-k}^{(i)f}d^{(p)f,++}_{-l}+k\g^{B(p)(i)}_{kl}\a_{-+,-l}^{(i)f}d^{(p)f,+-}_{-k}\right.\nn
&&\qquad\left. {}+\g_{R+-,kl}^{F(i)(p)}\a_{++,-l}^{(p)f}d^{(i)f,++}_{-k}+\g^{F(i)(p)}_{R+-,kl}\a_{-+,-l}^{(p)f}d^{(i)f,+-}_{-k}\right)|\chi(w_{1},w_{2})\rangle.
\label{Gplus2}
\eea

We now make the following convenient modifications to (\ref{Gplus2}):
\bea
\text{First Term}&:& k\leftrightarrow l\nn
&&\g^{B}_{kl}=\g^{B}_{lk}\nn
&&(i)\leftrightarrow (p)\nn\nn
\text{Fourth Term}&:&(i)\leftrightarrow (p).
\eea
This gives
\bea
0&=&-iC_{R+-}\sum_{(i),(p)}\sum_{k,l>0}\a_{++,-l}^{(p)f}d^{(i)f,++}_{-k}\left(k\g^{B(p)(i)}_{kl}+\g^{F(i)(p)}_{R+-,kl}\right)\nn
&&-iC_{R+-}\sum_{(i),(p)}\sum_{k,l>0}\a_{-+,-l}^{(i)f}d^{(p)f,+-}_{-k}\left(k\g^{B(p)(i)}_{kl}+\g^{F(p)(i)}_{R+-,kl}\right),
\eea
which implies
\bea
\g^{F(p)(i)}_{R+-,kl}&=&-k\g^{B(p)(i)}_{kl}\nn
\g^{F(p)(i)}_{R+-,kl}&=&\g^{F(i)(p)}_{R+-,kl},
\eea
for all $k,l>0$. The second was seen in (\ref{symmetryrequirements}).  The first is a manifestation of the supersymmetry in the theory.

We will note here that we were not required to map to a cover space beyond verifying that $G^+_{\dot A,0}\s_2^+$ vanishes. Since this vanishing should be a local property of the operators, independent of our winding configurations, we expect the relation proved here to extend to arbitrary twist configurations.

\end{document}